\documentclass[a4paper,11pt]{article}
\pdfoutput=1 
\newcommand{\be}{\begin{eqnarray}}
\newcommand{\ee}{\end{eqnarray}}
\newcommand{\nn}{\nonumber}
\newcommand{\nl}{\nonumber \\}
\newcommand{\pd}{\partial}

\usepackage{jheppub} 

\usepackage[T1]{fontenc} 
\usepackage{bm} 
\usepackage{subcaption}
\usepackage{mdframed}
\usepackage{enumitem}
\usepackage{multicol}
\usepackage{xcolor}
\usepackage{subcaption}
\usepackage{mathtools}
\usepackage{tikz-cd}
\usepackage{ulem}
\newmdenv[skipabove=6mm]{kotak}   

\newtheorem{definition}{Definition}

\title{\boldmath  Global flow structure and exact formal transseries of the Gubser flow in kinetic theory}

\date{}

\author[a]{Alireza Behtash,}
\author[a,b]{Syo Kamata,}
\author[a]{Mauricio Martinez,}
\author[a]{and Haosheng Shi}

\affiliation[a]{
	Department of Physics, North Carolina State University, Raleigh, NC 27695, USA}

\affiliation[b]{
	College of Physics and Communication Electronics, Jiangxi Normal University, Nanchang 330022, China
}

\emailAdd{abehtas@ncsu.edu}
\emailAdd{skamata11phys@gmail.com }
\emailAdd{mmarti11@ncsu.edu}
\emailAdd{hshi3@ncsu.edu}

\abstract{
In this work we introduce the generic conditions for the existence of a non-equilibrium attractor that is an invariant manifold determined by the long-wavelength modes of the physical system. We investigate the topological properties of the global flow structure of the Gubser flow for the Israel-Stewart theory and a kinetic model for the Boltzmann equation by employing Morse-Smale theory. We present a complete classification of the invariant submanifolds of the flow and determine all the possible flow lines connecting any pair of UV/IR fixed points. The formal transseries solutions to the Gubser dynamical system around the early-time (UV) and late-time (IR) fixed points are constructed and analyzed. It is proven that these solutions are purely perturbative (or power-law asymptotic) series with a finite radius of convergence. Based on these analyses, we find that Gubser-like expanding kinetic systems do not hydrodynamize owing to the failure of the hydrodynamization process which heavily relies on the classification of (non)hydrodynamic modes in the IR regime. This is in contrast to longitudinal boost-invariant plasmas where the asymptotic dynamics is described by a few terms of the hydrodynamic gradient expansion.  We finally compare our results for both Bjorken and Gubser conformal kinetic models.
}

\begin{document} 
	\maketitle
	\flushbottom

	\section{Introduction and summary}
	
	It is often argued in science and engineering textbooks that the regime of validity and applicability of hydrodynamics is linked to the existence of a hierarchy between the mean free path $\lambda_{m.f.p.}$ and the system size $L$. The existence of such a hierarchy implies that the gradients of the fluid hydrodynamic fields are small compared to their corresponding thermal equilibrium configurations. In a modern approach, this also gives rise to the interpretation of hydrodynamics as an effective field theory for the slowly varying long-wavelength modes of many-body systems in the form of a gradient expansion.
	
	This viewpoint, however, has been questioned recently in light of the phenomenological success of fluid dynamical models to describe rapidly expanding fireball of nuclear matter created in Ultra-relativistic Heavy Ion Collisions (cf. Refs.~\cite{Floris:2019klr,Li:2017qvf,Dusling:2015gta,Weller:2017tsr} and references therein). As a matter of fact, different toy models in the strong and weakly coupled regimes have hinted at a possibility of hydrodynamics being able to extend further to describe dynamical aspects of extremely non-thermal equilibrium configurations~\cite{Kurkela:2015qoa,Critelli:2017euk,Denicol:2014xca,Florkowski:2013lza,Florkowski:2013lya,Denicol:2014tha,Chesler:2009cy,Heller:2011ju,Florkowski:2017jnz,vanderSchee:2012qj,Chesler:2016ceu,Martinez:2010sc}. Furthermore, it was shown recently in  holographic systems~\cite{Grozdanov:2019kge,Grozdanov:2019uhi,Withers:2018srf} that the radius of convergence of the dispersion relations is orders of magnitude larger than the standard small-frequency limit. Similar conclusions were drawn in some kinetic theory models~\cite{Romatschke:2015gic,Kurkela:2017xis}. Therefore, we need a consistent fluid dynamics formulation to explain hydrodynamics far from equilibrium using first principles. 
	
	There are many unknown elements of this new formulation despite a few important lessons learned in the recent years. For instance, different numerical studies indicate that the equations of motion for a class of fluids undergoing Bjorken flow~\cite{Casalderrey-Solana:2017zyh,Strickland:2017kux,Florkowski:2017jnz,Romatschke:2017vte,Spalinski:2018mqg,Kurkela:2018xxd,Kurkela:2018oqw,Heller:2016rtz,Heller:2018qvh,Heller:2015dha,Strickland:2019hff,Strickland:2018ayk,Almaalol:2018jmz,Almaalol:2018ynx,Blaizot:2019scw,Blaizot:2017ucy,Blaizot:2017lht,Jaiswal:2019cju,Denicol:2019lio} can be dimensionally reduced to a single equation for the inverse Reynolds number $Re^{-1}$ as a function of the Knudsen number $Kn$~\footnote{For the Bjorken flow, the inverse Reynolds number is $\pi/p$ in which $\pi$ is the independent shear viscous component and $p$ is the equilibrium pressure, whereas the Knudsen number is $Kn\sim (\tau T)^{-1}$. We should also note that $Kn$ is sometimes denoted as $w$.}. Being an explicitly time-dependent (also known as ``nonautonomous'') dynamical system with an attracting IR fixed point~\footnote{We admit that the use of word fixed point in a time-dependent system is loose as it does not really exist for all times. However, at late times, we can take an open set localized at the equilibrium values for the configuration-space variables except time and consider that the flow remains oblivious to the time steps, making it a {\it limit} fixed point.} at late times $Kn \ll 1$, the model enjoys solutions that approach this fixed point which naturally merge before equilibriating. Roughly speaking, the set of points shared by the solutions to the model, aka ``flow lines'' in a dynamical system interpretation, is dubbed as {\it hydrodynamic attractor} or simply attractor. Since there are two UV fixed points in the system sourcing the flow of matter at early times $Kn\gg 1$, the main question is whether the attractor can be continued back to one of these fixed points~\footnote{It is possible for this continuation to not generate a unique critical line in case both the UV fixed points are ``saddle'', a fact that does not obviously hold in the configuration space of the Bjorken flow. Note that a critical line is really a complete flow line that basically lies on the boundary of basin of attraction~\cite{Behtash:2019txb}. }. In other words, is there a complete flow line initiated at a UV fixed point that connects to the hydrodynamic attractor? 
	
	Finding an answer to the question might ultimately boil down to considering the fast-slow decomposition of the geometry of distribution for a given system \cite{Gorban_2013,mckean_simple_1969}. Although a distribution obeying Boltzmann equation includes wider class of solutions in general, it can directly correspond to hydrodynamics through a lifting/projection process once a particular set of solutions are chosen.
	Above all else lies the fact that the existence of a slow manifold (or hydrodynamic attractor) is a necessary and sufficient indicator for hydrodynamization. However, the existence proof is a nontrivial problem for which one has to figure out both the global and local structures of the underlying dynamical system.
	
	A definite answer to this difficult problem can at least be given for certain kinetic models of expanding plasmas undergoing Bjorken flow. The longitudinal boost-invariant systems are often described in terms of the longitudinal proper time $\tau=\sqrt{(x^0)^2-(x^3)^2}$ and they present an unresolvable singularity in the Boltzmann equation at $\tau=0$~\cite{Behtash:2019txb}. It has been known for a while that certain interesting features of the hydrodynamic expansion of the Bjorken flow are better understood if we use the variable $w=\tau T$ ($T$ being the temperature)~\cite{Heller:2011ju}. However, in this new variable the UV structure of flow lines of the original dynamical system of ODEs is completely altered topologically~\footnote{The variable $w$ does not capture any UV information from the original system and rather serves as a toy model for the Bjorken flow in $\tau$. A simple pathology to think of is the fact that writing the equations in $w$ takes away the singularity at $\tau=0$ and transforms it into an unphysical singularity in another configuration-space component, the normalized shear viscous tensor. This makes $w$ a topologically forbidden coordinate change as far as the UV limit is concerned~\cite{Behtash:2019txb}. We have listed two other reasons at the end of Sec.~\ref{sec:conclusions}.}. This becomes a crucial matter if we are to study the UV completion of the flow lines. That is why it is sufficient to focus on the topology of the {\it invariant manifold} which contains all the stable flow lines that connect the physical information of the UV to that of IR. Thanks to the symmetries of the dynamical system, it is an extremely important point to press on that the topology of the invariant manifold has to survive any coordinate transformation respecting those symmetries, a requirement that is not satisfied for the $w$ variable.
	
	In a more general class of hydrodynamic theories, the search for solutions depends on the existence of an ``invariant manifold'' which substantiates many properties of the original kinetics. Since a reduced description of hydrodynamics only needs parametrization of microscopic variables projected as macroscopic fields ${\boldsymbol v}, p, \varepsilon$, a ``lifting procedure'' is required to acquire knowledge about a complete microscopic picture. This procedure is well-defined on a ``slow'' invariant manifold -- or simply slow manifold \footnote{In non-relativistic kinetic theory it is usually considered that if the Navier-Stokes limit is achievable, the slow manifold is then defined by this limit where long-wavelength modes attenuate at the some decay rate. The fast-slow decomposition connects the initial configurations of macroscopic fields (initial condition for ${\boldsymbol v}, p, \varepsilon$) to the hydrodynamization in the long-wavelength limit where the flow of hydrodynamics can be consistently defined from these macroscopic fields.}, where the hydrodynamic flow meets the kinetic motion of a fluid as the kinetics becomes well-behaved at a small Knudsen number and hydrodynamic field becomes bounded close to the local equilibrium  \cite{Gorban_2013,mckean_simple_1969,Tsumura:2011cj}. If the slow manifold exists, a complete understanding of hydrodynamics can be obtained based on these macroscopic fields. Therefore, an important task in this formulation is to find the slow manifold.
	
\begin{figure}[htpb]
		\centering
		\includegraphics[width=0.8\linewidth]{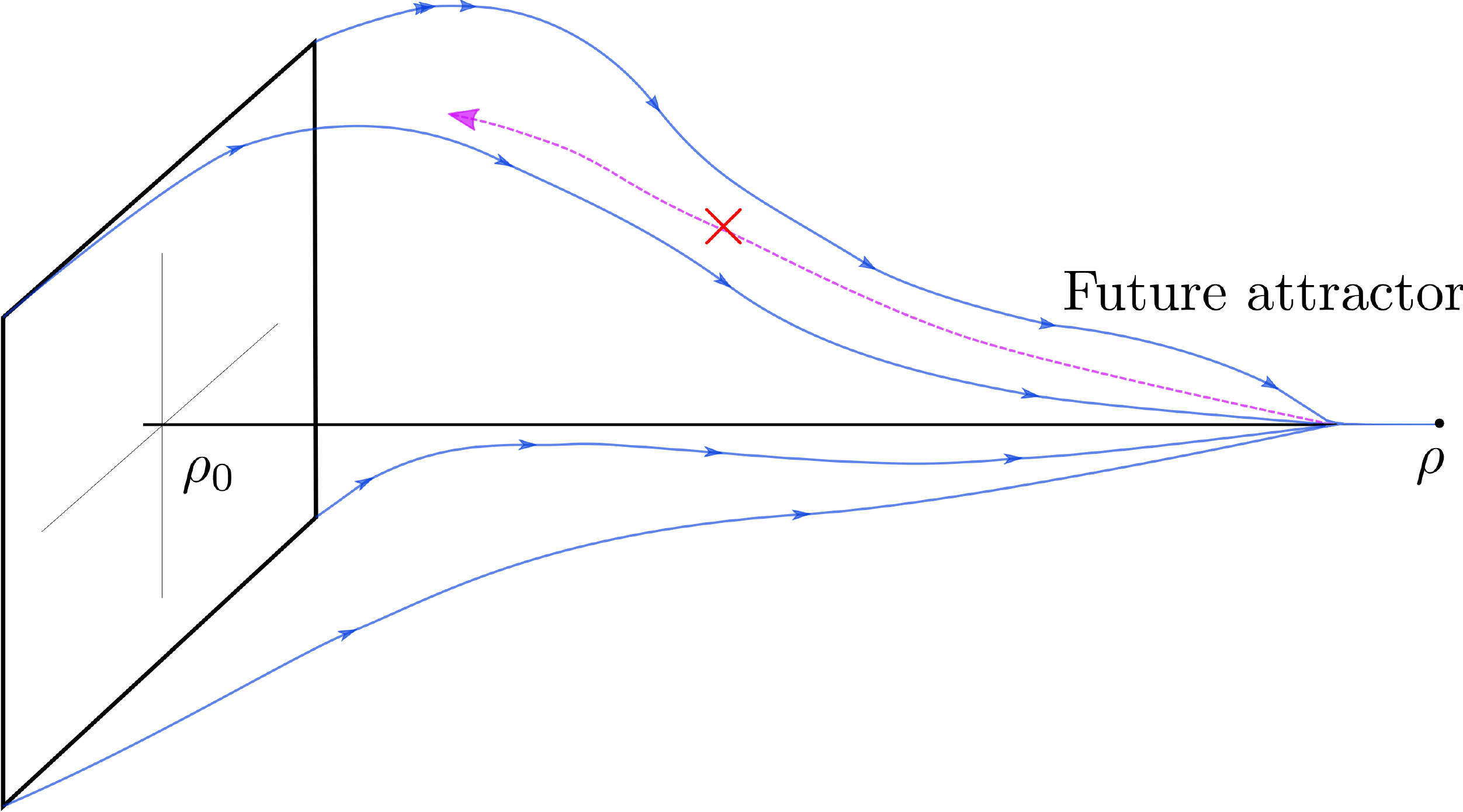} 
		\caption{Phase portrait of some nonautonomous system. At the point $\rho_0$ the equilibrium state in the future has not been formed yet. If we wait infinitely long enough, this state will start to form at $\rho$. As this is happening, the flow lines at late times become further dense around the equilibrium point. This dense set is the future (forward) attractor. Notice that there is no attraction in the past of $\rho$ due to the non-existence of an
			equilibrium state (fixed point), a statement that does hold in autonomous systems -- cf. RG flow space of field theories.}
		\label{fig:futureatt}
	\end{figure}

	Another misconception that we have to shed some light on is that in nonlinear nonautonomous systems, the dynamical system itself evolves with time so speaking of a global attractor that servers as a UV-IR complete flow line as in autonomous systems is not generally an option. Let us look at what this means with an example. Consider the Boltzmann equation with some collision kernel $\mathcal{C}[f(X(\tau), \tau)]$ \footnote{Normally, we should have included temperature in $f$, which is obviously ignored here, since it does not affect the conclusion.} which explicitly depends on time $\tau$ and some configuration-space variable $X$. Now, solving $\mathcal{C}[f]=0$ puts a condition on the distribution function. We can take this condition to be $f(X^*(\tau), \tau)=f_*$ where $f_*$ is a real constant.
	Here, $X^*(\tau)$ determines the zeros of the equation $f(X^*(\tau), \tau)-f_*=0$ and suppose for simplicity that there is only one zero and $\mathcal{C}[f(X(\tau))]$ is decreasing as we tend to this point. Normally, we would have considered this zero to be an attracting fixed point of the Boltzmann equation, but now obviously it is evolving with time. However, the evolution of $X(\tau)$ with appropriate initial values is going to get closer to $X^*(\tau_*)$ at every time step $0<\tau\le\tau_*$. If we repeat the same exercise of obtaining the evolution of individual variables until $\tau_*$ for a new $\tau_*$, we can observe a line across all values of $\tau_*$. This line is deemed as an ``attractor solution'' since it naturally attracts the points falling in the basin of attraction of $X^*(\tau)$ until every time step $\tau_*$, even though as we can see, it is nothing but the trace of the zero of collision kernel across all times which really becomes the thermal equilibrium of the system at $\tau_*\rightarrow\infty$. In other words, the system did not know about its equilibrium state in the past since there was none to begin with. Only it is at $\tau_*\rightarrow\infty$ that the system reaches a fixed point $X^*(\tau_*)$ and an equilibrium distribution $f_*$. Therefore, we need to really talk about attractors in the past or future of an evolving fluid with explicitly time-dependent dynamics; see Fig.~\ref{fig:futureatt}.
	
	The analysis of higher dimensional fluids such as the Gubser flow~\cite{Behtash:2017wqg,Denicol:2018pak} as well as less symmetric flow profiles in $2+1$D~\cite{Romatschke:2017acs} and $3+1$D~\cite{Kurkela:2019set} indicate that there are attractors for these PDEs that might be subset of complete flow lines. The analysis of attractors for higher dimensional systems is more involved than the 1D longitudinal boost-invariant case so the original approach of the authors of Ref.~\cite{Heller:2015dha} falls short to describe them. For instance, for the Gubser flow, two of the current authors showed by studying that flow as a dynamical system and performing stability analysis of IS theory~\cite{Behtash:2017wqg} that the form of attractor is not described by the asymptotic perturbative gradient expansion in terms of the Knudsen number. Although this example was just a simple exercise to show the power of dynamical systems in lower spatial dimensions, this approach in essence is way more generic and could apply to fluids expanding in higher dimensions for probing not only the attractors, but also interesting phenomena such as the existence of saddle-type steady states in the mid-range, fluid turbulence, chaos, etc.
	
	In this work, we investigate and generalize our previous findings~\cite{Behtash:2017wqg} and focus on constructing all the possible links between UV and IR points using topological properties of the Gubser flow as a nonautonomous dynamical system based on the Morse-Smale decomposition and the transseries analysis. In practice, it is possible to describe the story using just the Morse decomposition, in which
	case we should focus on the Morse-Lyapunov functions for describing the flows. So for us, the two decompositions are interchangeable
	even though the demonstrations will be done without referring to a specific Morse function.
	
	To start things off, we consider the global structure by giving an account of the stability analysis of the fixed points ~\cite{kloeden2011nonautonomous,caraballo2017applied} and define rigorously what exactly an attractor is in a nonautonomous system. This comes handy when we discuss different types of attractors in this system. We consider the time evolution of flow lines (configuration-space trajectories) attached to individual fixed points portrayed in lower dimensions in the space of moments of distribution function and temperature, and reveal the global structure from the topology of its invariant set and constituent invariant subsets. It is shown that there exists two bounded invariant submanifolds in the configuration space and one of them has a nontrivial flow structure in the sense that it contains a local future attractor, but the system lacks a local past attractor. As for the analytic solutions to the evolution equations, we construct a formal transseries solution by expanding around the stable IR fixed point and
	show the asymptotic behavior of hydrodynamical quantities and the Boltzmann distribution function. Furthermore, we prove that the formal transseries is a {\it convergent series} with a finite radius of convergence \footnote{In an earlier paper \cite{Behtash:2017wqg}, we did only consider a few terms in the naive asymptotic expansion around the stable IR fixed point and concluded that it was divergent. In this paper, we show that in fact had we continued further, the expansion would have turned out to be convergent eventually.}, and hence why the distribution does not converge to a thermal equilibrium. Indeed, one cannot find a well-defined hydrodynamic description of Gubser flow due to the fact that the Knudsen number is large and a slow manifold does not exist. This is backed by the failure of Chapman-Enskog expansion, which stems from the nonvanishing derivative of distribution at late times, implying that the system still undergoes expansion near the local equilibrium. We conclude by giving final remarks.
	
	\section{Stability analysis of conformal Israel-Stewart theory}
	
	We start our discussion of the global flow structure and stability analysis for the Gubser flow by considering the Israel-Stewart (IS) theory~\cite{Israel:1979wp}. As a warm-up exercise, we use this hydrodynamic truncation scheme to get things started and introduce a set of mathematical tools to analyze the structure of the solutions near its asymptotic early-time (UV) and late-time (IR) fixed points. Some of the results presented in this section were partially derived in a previous publication~\cite{Behtash:2017wqg}. We are going to build some aspects of our original approach into this paper and extend our previous findings. The generalization of the methods described in this section to the case of the relativistic Boltzmann equation within the relaxation time approximation (RTA) is presented in forthcoming sections. 
	
	The Gubser flow describes a conformal fluid undergoing longitudinally boost-invariant and azimuthally symmetric expansion on the transverse plane in Minkowski space~\cite{Gubser:2010ze,Gubser:2010ui}. {\color{blue} } The symmetry group of this flow, $SO(3)_q\times  SO(1,1)\times  \mathbb{Z}_2$ is then manifestly inherited from the symmetries of $dS_3\times \mathbb{R}$ spacetime and thus, the analysis of dissipative equations is more conveniently described on this manifold. The coordinates, the metric $g_{\mu\nu}$, and the flow velocity profile in this curved space are respectively given by
	\be
	&& x^{\mu}=(\rho,\theta,\phi,\varsigma), \qquad \rho,\varsigma \in {\mathbb R}, \qquad \theta \in [0,\pi], \qquad \phi \in [0,2 \pi), \\
	&& g_{\mu \nu} = {\rm diag}(1,-\cosh^2 \rho,-\cosh^2 \rho \sin^2 \theta,-1),\\
	&&u^\mu=(1,0,0,0)\,.
	\ee
	From the conservation law and the IS theory, the equations of motion for the effective temperature $T$ and the normalized shear viscous component $\overline{\pi} :={\pi}^{\varsigma\varsigma} /({\epsilon}+p)$~\footnote{Here, $\pi^{\varsigma\varsigma}$ is the only independent component of the shear viscous tensor, $\epsilon$ is the energy density, and $p$ is the equilibrium pressure. In our previous works~\cite{Behtash:2017wqg,Behtash:2018moe,Behtash:2019txb}, we called $\overline{\pi}$ the effective normalized shear viscous tensor. We will stick to this name throughout the present paper as well.} are derived to be~\cite{Marrochio:2013wla}
	\begin{subequations}
		\label{eq:ISodes}
		\begin{align}
		\frac{d T}{d \rho} &=  \frac{T}{3} \left( \bar{\pi}-2 \right) \tanh \rho, \label{eq:IS_T} \\
		\frac{d \bar{\pi}}{d \rho} &= -\frac{4}{3}  \left( \bar{\pi}^2-\frac{1}{5} \right) \tanh \rho - \frac{T \bar{\pi}}{c}. \label{eq:IS_pi}
		\end{align}
	\end{subequations}
	where $c>0$ is an arbitrary dimensionless real scale that relates the shear relaxation time scale $\tau_\pi$ and the ratio of shear viscosity over entropy $\eta/s$, which in the conformal case is simply $\tau_\pi T = c\,\eta/s$.  Eqs.~\eqref{eq:ISodes} constitute what is called in the math literature a \textit{nonautonomous dynamical system} since the de Sitter time variable $\rho$ appears explicitly on the r.h.s. through the expansion rate $D_\mu u^\mu=\tanh\rho$~\footnote{
	The nonautonomous dynamical system is a group of ODEs which depend explicitly on time, i.e., 
\begin{equation}
\frac{d {\bf A}}{dt}=S(t,{\bf A})\,,
\end{equation}
being $S(t,{\bf A})$ an arbitrary function of the vector field ${\bf A}$ and the time variable $t$. If the RHS of the previous generic equation does not have an explicit time dependence the system is called autonomous. }.
	
	In general, a nonautonomous system has some subtleties when it comes to defining {}and analyzing its dynamical equilibrium points, i.e. fixed points, and their stability. For instance, Eqs. \eqref{eq:ISodes} do not have a fixed point on the finite domain unless at $\rho\rightarrow\infty$. This is because 
	the r.h.s. of Eqs. \eqref{eq:ISodes} includes $\rho$ explicitly so only for infinitely large $|\rho|$,
	the configuration-space variables become independent of $\rho$, thus constant under flow time translations. Since the time dependence of the flow is explicit,  the said statement means that the fixed points of a nonautonomous systems are only asymptotically achieved and therefore live on the boundary of configuration space \footnote{For this reason, if the time direction is non-compact, it is more appropriate to use the term ``fixed points at infinity'' when referring to the fixed points of nonautonomous systems; see for instance \cite{berger2005one}. It should be reminded that throughout this paper, we will assume the existence of fixed points for a fixed time interval \cite{layek2015introduction}.}. Moreover, in nonautonomous systems, a smooth change in the flow parameter can induce topological changes such as bifurcation that are probed in time-independent systems by tuning a control parameter. Consequently, the knowledge of local behavior is not adequate to determine the UV completion of the solution. A holistic comprehension of the flow is needed, facilitated by the global analysis of its topology. There are other differences that we will touch on as we continue.
	
	\subsection{Flow structure}
	In this subsection, we classify all the fixed points of the IS theory in both UV and IR limits by calculating their
	instability index in the extended configuration space to lay out the basic local structure of the flow. We use the skew product flow formalism of this nonautonomous system to portray both the local structure of the configuration space $(T(\rho), \bar \pi(\rho))$ and its time-dependent change in the global picture. Following this construction and Morse decomposition, we show all the possible flow lines connecting any pair of UV/IR fixed points and discuss the existence or nonexistence of pullback attractors.
	
	\subsubsection{Fixed point and stability}
	
	This subsection overlaps with previous findings~\cite{Behtash:2017wqg} and recaps some of those results. We consider the analysis of fixed points in the UV and IR for Eqs.~\eqref{eq:ISodes} determined by finding the zeros of the expressions on the right-hand sides. This basically amounts to solving for points at which $d \bullet / d \rho = 0$ and taking the limit $\rho \rightarrow \mp \infty$ in the solutions. 
	The original nonautonomous dynamical system~\eqref{eq:ISodes} can be made autonomous by defining a flow time variable $\tau=\tanh\rho$. The resulting autonomous system reads~\footnote{In the language of dynamical systems, the time-independent ODEs are often called \textit{master-slave system} \cite{Kloeden:2011,caraballo2017applied}.} 
	\be
	\label{eq:ISaut}
	&&\frac{d T}{d \rho} = \frac{T}{3} (\bar{\pi} - 2 ) \tau =: F_T (X), \label{eq:IS_T_3d} \\
	&&\frac{d \bar{\pi}}{d \rho} = -\frac{4}{3} \left(  \bar{\pi}^2 - \frac{1}{5} \right) \tau - \frac{T \bar{\pi}}{c} =: F_{\bar{\pi}}(X), \label{eq:IS_pi_3d} \\
	&&\frac{d \tau}{d \rho} = 1-\tau^2 =: F_{\tau}(X), \label{eq:IS_tau_3d}
	\ee
	where we have defined $X:= (T,\bar{\pi},\tau)$. One of the major advantages of performing this transformation is that the fixed points are no longer at infinity since $\tau=\mp 1$~\cite{Behtash:2017wqg}. As a result, fixed points on the finite-time boundary naturally unfold the interpretation of equilibria in a hydrodynamic system. The fixed points $X^c$ defined by $F_i(X)=0$ for any index $i$ are given by
	\be
	X^c = 
	\begin{cases}
		(0,+1/\sqrt{5},-1)=:\Sigma_{\rm U(A)}  \\
		(0,+1/\sqrt{5},+1)=:\Sigma_{\rm I(A)}  \\
		(0,-1/\sqrt{5},-1)=:\Sigma_{\rm U(B)} \\
		(0,-1/\sqrt{5},+1)=:\Sigma_{\rm I(B)} \\
		(+38c/15,2,-1) =: \Sigma_{\rm U(C)} \\
		(-38c/15,2,+1) =: \Sigma_{\rm I(C)} \\
	\end{cases} \label{eq:crit_point_IS}
	\ee
	where $\Sigma_{\rm U(\bullet)}$ and $\Sigma_{\rm I(\bullet)}$ denote the UV ($\tau=-1$) and IR ($\tau=1$) fixed points respectively.
	We define the fixed point set as $\Sigma_a=\{\Sigma_{\rm U(A)},\Sigma_{\rm I(A)},\dots,\Sigma_{\rm U(C)},\Sigma_{\rm I(C)}\}$ where $a=\{{\rm U(A)},{\rm I(A)}, \dots \}$.
	
	The stability for each fixed point can be determined by slightly perturbing the solution in the vicinity of it. This leads to a linearized form of the ODEs which captures whether a small deviation around the fixed point ${\bf X} \rightarrow  {\bf X}^c\,+\,\delta {\bf X}$ is attractive (a sink), repulsive (a source), a combination of both (a saddle), etc. In an autonomous system, an equilibrium is classified by the dimension of attraction or repulsion, which is probed by the direction where the deviation of a system is uniformly bounded or the opposite. The importance of these qualities from a physical standpoint is that we can label the points at which the flow lines are starting to diverge from or approach ultimately in the UV and IR limit by identifying these points as either stable, marginally stable, unstable, etc. For example, hydrodynamic equilibrium is always a stable point, which means that every flow line in the vicinity of this point will be attracted to it. The generic form of a linearized ODE reads 
	\be
	\frac{d \delta {X}}{d \rho} = J^{a} \delta {X},
	\ee
	where $J^{a}$ is called Jacobian matrix. For the autonomous system~\eqref{eq:IS_T_3d}-\eqref{eq:IS_tau_3d}, $J^{a}$ is defined by
	\be
	J^{a}_{ij} := \left. \frac{\pd F_{i}}{\pd X_j} \right|_{X\rightarrow \Sigma_a} = 
	\left.
	\begin{pmatrix}
		\frac{(\bar{\pi}-2) \tau}{3} & \frac{T \tau}{3} & \frac{T(\bar{\pi}-2)}{3} \\
		-\frac{ \bar{\pi}}{c} &  -\frac{8 \bar{\pi}}{3} - \frac{T}{c} &  -\frac{4}{3} \left( \bar{\pi}^2 - \frac{1}{5} \right) \\
		0 & 0 & -2 \tau
	\end{pmatrix} \right|_{X \rightarrow \Sigma_{a}}.
	\ee
	One can easily obtain the eigenvalue of $J^a$ as
	\be
	\label{eq:eigenJ}
	{\rm Eigen} (J^a) &=& {\rm diag}(b^a_1,b^a_2,b^a_3) \nl
	&=&
	\begin{cases}
		\left( \pm \frac{1}{3} \left( 2 - \frac{1}{\sqrt{5}} \right), \,  \pm \frac{8}{3 \sqrt{5}} ,\, \pm 2 \right) & \mbox{ for } \Sigma_{\rm U,I(A)}  \\
		\left( \pm \frac{1}{3} \left( 2 + \frac{1}{\sqrt{5}} \right), \, \mp \frac{8}{3 \sqrt{5}} ,\, \pm 2 \right) & \mbox{ for } \Sigma_{\rm U,I(B)}  \\
		\left( \pm \frac{1}{15} \left(\sqrt{821} +21 \right), \mp \frac{1}{15} \left( \sqrt{821}-21 \right) ,\pm 2 \right) & \mbox{ for } \Sigma_{\rm U,I(C)}.  
	\end{cases}
	\ee
	In order to measure the stability in the vicinity of a fixed point, we define the {\it instability index}~\cite{thompson2002nonlinear} \footnote{From the perspective of Morse theory, instability index is identical to the Morse index of a critical point, which is 
		the dimension of the unstable manifold of the critical point over which the Hessian is negative-definite. For example, a flow connected to a fixed point with instability index of one should move away along the direction of the corresponding eigenvector of $J^a$, which is equivalent to saying that we have a critical point of index one around which the associated Morse function is concave down, thus a 1-dimensional unstable manifold.}
	\be
	{\rm Ind}[J^a] := {\rm Ind}[J^a] := \sum_{i=1}^3 H(\Re\,b^a_i)\,,
	\label{eq:Index_def}
	\ee
	where $H(x)$ is the Heaviside function. For the eigenvalues~\eqref{eq:eigenJ} one obtains the following indexes for each fixed point
	\be
	\label{eq:Ind}
	{\rm Ind}[J^{\rm U(A)}] = 3, \qquad {\rm Ind}[J^{\rm I(A)}] = 0, \label{eq:IndA} \\
	{\rm Ind}[J^{\rm U(B)}] = 2, \qquad {\rm Ind}[J^{\rm I(B)}] = 1, \label{eq:IndB} \\
	{\rm Ind}[J^{\rm U(C)}] = 2, \qquad {\rm Ind}[J^{\rm I(C)}] = 1. \label{eq:IndC}
	\ee
	According to the index (\ref{eq:Index_def}), one can see that $\Sigma_{\rm U(A)}$ and $\Sigma_{\rm I(A)}$ are a source and a sink, respectively, and the others are saddles. The behavior of the flow lines in the vicinity of a particular fixed point in this case is determined by the eigenvalues $b^a$ of the matrix $J^a$. In the three dimensional configuration space $X$, there are 8 possible fixed points:
	\begin{itemize}
		\item One source if all the eigenvalues are positive;
		\item One sink if none of the eigenvalues are positive;
		\item One index-1 saddle point if all eigenvalues are real with only one being positive;
		\item One index-2 saddle point if all eigenvalues are real with only two being negative;
		\item One spiral attractor if there is one real eigenvalue and one complex-conjugate pair with all having negative real parts;
		\item One spiral repeller if there is one real eigenvalue and one complex-conjugate pair which have positive real parts;
		\item One index-1 spiral saddle point with one positive real eigenvalue and two complex conjugate eigenvalues with negative real parts;
		\item One index-2 spiral saddle point with one negative real eigenvalue and two complex conjugate eigenvalues with positive real parts.
	\end{itemize}
	The value of the eigenvalues is directly linked with the behavior of the solutions of the linearized equation around the fixed point. For example, for $\Sigma_{\rm U,I(A)}$ one finds that 
	\be
	\label{eq:linsolsIS}
	&& \delta T(\rho)
	= \sigma_1 e^{\pm \frac{1}{3}(2-\frac{1}{\sqrt{5}})  \rho}  = \sigma_1 z_{\mp}^{-\frac{1}{6}(2-\frac{1}{\sqrt{5}}) }
	, \label{eq:IS_T_sle} \\
	&& \delta \bar{\pi}(\rho)
	= \frac{\sigma_1  e^{\pm \frac{1}{3}(2-\frac{1}{\sqrt{5}})  \rho} + \sigma_2 e^{ \pm \frac{8}{3 \sqrt{5}} \rho}}{(9 \sqrt{5}-10)c}
	=  \frac{\sigma_1  z_{\mp}^{-\frac{1}{6}(2-\frac{1}{\sqrt{5}})} + \sigma_2 z_{\mp}^{- \frac{4}{3 \sqrt{5}}}}{(9 \sqrt{5}-10)c}
	, \label{eq:IS_pi_sle} \\
	&& \delta \tau(\rho) = \sigma_{3} e^{\pm 2 \rho} = \sigma_{3} z_{\mp}^{-1},
	\ee
	where $z_{\pm}=e^{\pm 2 \rho}$, $\sigma_i$ is an integration constant and the Lyapunov exponents correspond to the eigenvalues governing the damping and/or growth of the perturbation close to the fixed points. It is important to emphasize that on the one hand, the solutions of the linearized ODEs give the transmonomial of the formal transseries, which are usually identified as a {\it non-perturbative term} in the sense that exponentially suppressed terms do not admit a perturbative expansion~\cite{Costin:2008,Behtash:2019txb,Behtash:2018moe,Basar:2015ava}. Since the linearized solutions~\eqref{eq:linsolsIS} do not have the typical exponential transmonomials when written in terms of $z_{\mp}$, the solutions are deemed merely perturbative even though there might be power-law corrections. On the other hand, the asymptotics of IS theory describes a far-from-equilibrium state due to the curvature of $dS_3\times \mathbb{R}$ when $\rho\to\pm\infty$, and it thus cannot have a hydrodynamic gradient expansion. Another way to see this is that the solutions of IS theory for the Gubser flow do not scale asymptotically with the Knudsen number $Kn\sim \tanh\rho/T(\rho)$. In other words, the demise of gradient expansion is a direct result of the break of Knudsen number in the presence of nonvanishing curvature in the IR limit. In the next subsection, we will explain why the expansion parameter for the formal transseries must be $z_{\mp} = e^{\mp 2\rho}$ \footnote{This expansion is asymptotic and does not continue to small $|\rho|$ due to the fact that $z_{\mp}\rightarrow1$ as $\rho\rightarrow0$, meaning that a transseries solution
		does not exist across $\rho \in \mathbb{R}$ connected to the hydrodynamic equilibrium $\Sigma_{\rm I(A)}$ if the expansion parameter is just $z_{\mp}$. We prove the existence or (non-existence) of solutions between any possible pair of UV/IR fixed points systematically using topological arguments and construct transseries solutions in the next two subsections.}.
	
	\subsubsection{Global structure} \label{sec:global_IS}
\begin{figure}
		\centering
		\includegraphics[width=1.0\linewidth]{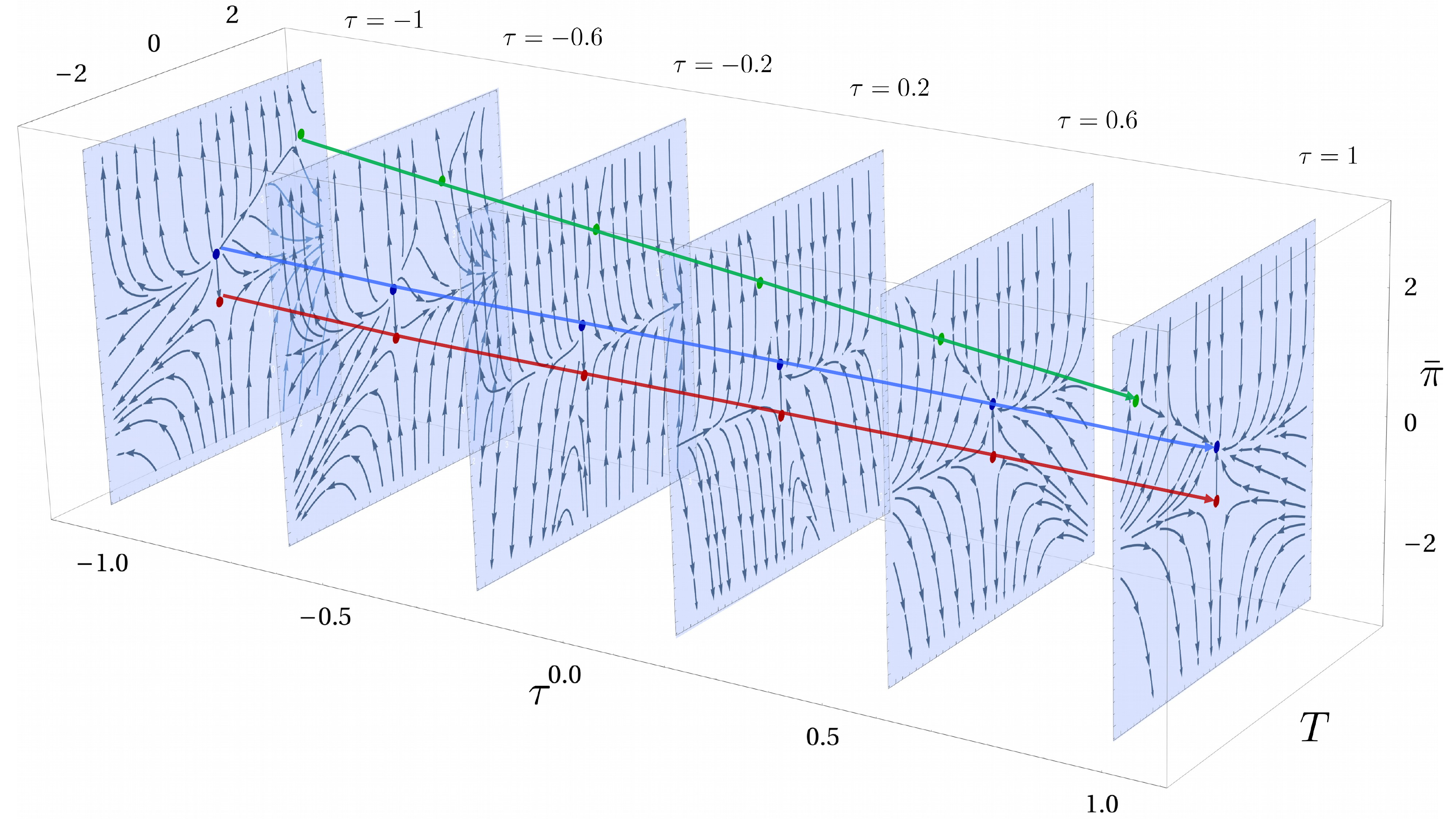}
		\caption
		{Flow structure on the $\tau$-fiber.
			The colored points denote $\Sigma_{\rm A}(\tau)$(Blue), $\Sigma_{\rm B}(\tau)$(Red), and $\Sigma_{\rm C}(\tau)$(Green) given by Eq.~(\ref{eq:core1}).
		}
		\label{fig:flow_IS1}
	\end{figure}

The nonautonomous system is a group of ODEs which depend explicitly on time.
The most distinguishing aspect of this type of system compared to an autonomous system is that in the latter, a flow is invariant under time translations, whereas in the former, it does care about  when the flow was initiated. In other words, an explicitly time-dependent dynamical system does entail a time memory for every flow line on the contrary to an autonomous system in which the flow dynamics does not keep track of the initial time at which we considered the flow. This far-reaching effect changes the way to survey a nonautonomous system as its flow line is contigent on when the history is defined. In contrast to an autonomous system where a flow propagating with the same period of time is identical on a manifold, the topological space of a nonautonomous system identifies a flow by both of its initial time and its evolution $\Delta\rho$ treated as a parameter. Therefore, a flow which is sufficiently well-characterized on a time-independent smooth manifold of configuration-space variables, needs to be generalized to a smooth {\it fiber bundle}. This basically makes the space of variables in the configuration space a spacelike hypersurface, e.g., fiber or transverse manifold, that evolves over the time direction -- also called base manifold (see Fig.~\ref{fig:flow_IS1})~\footnote{For the reason that there is an evolution of the spacelike hypersurfaces across the flow itself, the name ``process'' has been given to such a flow line in math literature,
		which with a slight abuse of language will be avoided here.}. Additionally, every spacelike hypersurface is labeled by the time step it is retrieved at. This is reminiscent of a spacelike hypesurface in relativity where the normal vector determines the time state throughout the hypersurface. Hence, time direction indeed becomes part of the manifold over which we let the flow lines be defined in a nonautonomous dynamical system such as Gubser flow. For the purpose of solving the whole system, it is then of utmost importance to try to get a handle on both global and local structure of the flow lines over this fibrated manifold.
	
	In order to understand the global structure, we employ \textit{skew product flow formalism} (SPFF) \cite{Kloeden:2011,caraballo2017applied}.
	This approach will require us to first construct a vector bundle with the fiber being the configuration space itself and a base space of the flow time as stated earlier, then introduce the flow lines on both spaces independently.
	The original nonautonomous dynamical system is realized using flows on the total space by taking an appropriate flow equation on the base space.
	If we fix the time, the resulting hypersurface (fiber) gives a dynamical system on the fiber space, which is a projection of the flows of the IS equations~\eqref{eq:ISodes} defined on the total system onto the fiber space at fixed time. This means that on every fiber, time acts as a fixed parameter so we can consider a parameter-dependent flow structure on the fiber space and study how it transforms under any variation of the time parameter -- that is not a variable anymore. One interesting phenomenon that could possibly happen is bifurcation, namely whether or not the flow hits a ``turn-off'' where it could go either way. This provides a powerful method for understanding the entire structure of original nonautonomous system through the lens of an autonomous system \cite{Behtash:2017wqg}.
	
	Let us define a trivial vector bundle $\Theta=({\cal M},\pi,B,F)$ for the nonautonomous dynamical system at hand.
	We define \footnote{
		We normally take the time direction to represent the base space in, for example, field theory applications,
		but in this analysis we assume $\tau$ is a new time variable and rather take the de Sitter time $\rho$ to play the role of an affine parameter.
	}
	\be
	\mbox{Total space} &:& {\cal M} = B \times F, \\
	\mbox{Projection} &:& {\pi} : {\cal M} \rightarrow B, \\
	\mbox{Base space} &:& B = I := (-1,+1) \ni \tau, \\
	\mbox{Fiber space} &:& F = {\mathbb R}^2 \ni {\bf X} = (T,\bar{\pi}).
	\ee
	Total space is sometimes called the extended configuration space or extended phase space. From a technical point of view, a fiber of a map $f:X\rightarrow Y$ is the preimage of an element $y \in Y$. Similarly, to define a $\tau$-fiber of the projection map $\pi$ defined above, we view ${\cal M}_{\tau}$ as the preimage of $\tau \in B$, that is $\pi^{-1} (\tau) =: \{ \tau \} \times {\cal M}_{\tau}$.
	Then, we define the nonautonomous dynamical system by the SPFF \cite{Kloeden:2011,caraballo2017applied} as follows
	\begin{kotak}
		\begin{definition}
			{\rm
				A \textit{nonautonomous dynamical system} $(\theta,\phi)$ is defined by a continuous cocycole mapping $\phi$ on the fiber space $F$ which is driven by an autonomous dynamical system $\phi$ acting on the base space $B$ and time set ${\mathbb T}={\mathbb R} \ni \lambda$.
				The dynamical system $\theta$ on $B$ is a group of homeomorphisms $\theta_{\lambda \in {\mathbb T}}$ with the properties that
				\be
				&& \theta_{0}(\tau_0) = \tau_0 \quad \mbox{for any } \tau_0 \in B, \\
				&& \theta_{\lambda_1 + \lambda_2 }(\tau_0) = \theta_{\lambda_1} (\theta_{\lambda_2}(\tau_0)) \quad \mbox{for any } \lambda_1,\lambda_2 \in {\mathbb T}, 
				\ee 
				and the cocycle mapping $\phi:{\mathbb T}^{+}_{0} \times B \times F \rightarrow F$ satisfies
				\be
				&& \phi(0,\tau_0,{\bf X}_0) = {\bf X}_0 \quad \mbox{for any } (\tau_0,{\bf X}_0) \in {\cal M}, \\
				&& \phi(\lambda_1 + \lambda_2,\tau_0,{\bf X}_0) = \phi(\lambda_1,\theta_{\lambda_2}(\tau_0), \phi(\lambda_2 , \tau_0,{\bf X}_0)) \nl
				&& \qquad \qquad \qquad \qquad \mbox{for any } \lambda_1,\lambda_2 \in {\mathbb T}^{+}_{0}, \ (\tau_0,{\bf X}_0) \in {\cal M}.
				\ee
			}
		\end{definition}
	\end{kotak}

	In this definition, the cocycle mapping \footnote{Cocycle is merely a tool to reassure that the flow has a group structure. This equips us with the ability to provide existence proofs for the flows between UV/IR fixed points based on topological arguments that depend on this group structure.} has a semi-group structure, but replacing ${\mathbb T}^+_0 \rightarrow {\mathbb T}$ yields a group instead since the system is invertible along the time direction. Namely, under time translations $\tau\rightarrow\tau + \tau_0$, every positive time $\tau>0$ is mapped to an inverse time represented by $-\tau<0$ and vice versa. ${\cal M}$ is called a \textit{nonautonomous set}.
	If there exists a submanifold ${\cal S} \subset {\cal M}$ such that $(\theta_{\lambda}(\tau), \phi(\lambda,{\cal S}))={\cal S}$ for any $\lambda \in {\mathbb T}$, we call $\cal S$ an \textit{invariant (sub)manifold of the flow.} The search for a solution of IS theory can be converted to the  discovery of a corresponding invariant manifold. Strictly speaking, the IS theory is only invertible in the invariant manifold because it does not intersect any other submanifold. 
	
	As explained in the introduction, the explicit time dependence does not allow for the dynamical equilibrium state to form unless the flow continues as time tends to infinity. This means that we should distinguish between the future and past domain of $\cal M$. Let us define a subspace given by ${\cal M}_{-(+)}:= \bigcup_{\tau \le 0 (\ge 0)} \{ \tau\} \times {\cal M}_{\tau}$ as the past(future) domain of the dynamical system.
	
	The dynamical system $(\theta,\phi)$ can be identical to the solution of the ODEs (\ref{eq:IS_T_3d})-(\ref{eq:IS_tau_3d}) for a given initial condition $(\tau_0,{\bf X}_0)$ at $\rho = \rho_0$  as $\theta_{0}(\tau_0):=\tau(\rho_0) = \tanh \rho_0, \phi(0,\tau_0,{\bf X}_0) := {\bf X}(\rho_0)$.
	Moreover, the flow parameter $\lambda$ and the flow time $\rho$ can be related to each others as $\theta_\lambda(\tau_0)=\tanh (\lambda+ {\rm arctanh} \, \tau_0)$, hence, $\rho = \lambda+ {\rm arctanh} \, \tau_0$.
	By setting an initial condition $(\tau_0,{\bf X}_0) \in {\cal M}$,
	the dynamical system $(\theta,\phi)$ gives a one-dimensional orbit on the vector bundle $\gamma_{\tau_0}({\bf X}_0) \subset {\cal M}$, which is a section determined by the ODEs, as 
	\be
	&& \gamma_{\tau_0}({\bf X}_0)= \{ (\tau,{\bf X}) \in \gamma   \, | \,  \gamma = \bigcup_{\lambda \in {\mathbb T}} (\theta_{\lambda}(\tau_0),\phi(\lambda,\tau_{0},{\bf X}_0)) \}.
	\ee
	It is notable that  due to the uniqueness theorem, all the flows are distinguished only by the initial condition ${\bf X}_0$ once $\tau_0$ is fixed.
	This property is not dependent on $\tau_0$ in the autonomous flows, tying the fate of flow lines solely to the initial condition ${\bf X}_0$.
	
	We would like to investigate the flow structure on the $\tau$-fiber ${\cal M}_{\tau}$ and its $\tau$-dependence.
	Instead of Eq.~\eqref{eq:IS_tau_3d}, if one takes $\theta_\lambda(\tau)$ to be an identity mapping, i.e. $\theta_{\lambda}(\tau) = \tau$ 
	for any $\lambda \in {\mathbb T}$, the resulting flows will live be on the $\tau$-fiber ${\cal M}_{\tau}$.
	What this means is that a flow on ${\cal M}_{\tau}$ determined by Eqs. (\ref{eq:IS_T_3d}) and (\ref{eq:IS_pi_3d}) is now treated as if
	$\tau$ is a free control parameter. In this sense, the projection of flow structure is solely determined by $\tau$ whilst the complete solution is determined by flow time $\rho$ as well, i.e., 
	\be
	&&\frac{d T}{d \rho} = \frac{T}{3} (\bar{\pi} - 2 ) \tau =: F_T ({\bf X};\tau), \label{eq:IS_T_2d_fiber} \\
	&&\frac{d \bar{\pi}}{d \rho} = -\frac{4}{3} \left(  \bar{\pi}^2 - \frac{1}{5} \right) \tau - \frac{T \bar{\pi}}{c} =: F_{\bar{\pi}}({\bf X};\tau). \label{eq:IS_pi_2d_fiber}
	\ee
	The solution depends on both $\rho$ and $\tau$, so it is written as ${\bf X}={\bf X}(\rho;\tau)$. Note that for a given initial condition ${\bf X}_0(\tau):={\bf X}(\tau,\rho_0)$, $\rho$ relates to 
	some $\lambda\in {\mathbb T}$ via a translation $\rho=\lambda +\rho_0$.
	
	The fixed points on the $\tau$-fiber are determined from Eqs. (\ref{eq:IS_T_3d}) and (\ref{eq:IS_pi_3d}) with $d T/d\rho=d \bar{\pi}/d \rho = 0$ to be
	\be
	&& \Sigma_{\rm A}(\tau) = \left( 0, + \frac{1}{\sqrt{5}}\right), \quad  \Sigma_{\rm B}(\tau) = \left( 0, - \frac{1}{\sqrt{5}}\right), \quad  \Sigma_{\rm C}(\tau) = \left( -\frac{38 c \tau}{15} , 2 \right). \label{eq:core1}
	\ee
In the above formulation, the fixed point is contingent on $\tau$ alone regardless of how the flow is initialized by specifying the relation of $\rho$ and $\tau$. This approach helps to determine the fixed points in a nonautonomous system by using the technique of an autonomous system. As a result, the equilibria of hydrodynamics is related to the fixed points at the boundary of $\tau$. One can immediately find that the solutions (\ref{eq:core1}) converge to the fixed points given by Eq. (\ref{eq:crit_point_IS}), i.e., $\lim_{\tau \rightarrow -1} \Sigma_{\bullet}(\tau) = \Sigma_{\rm U(\bullet)}$ and $\lim_{\tau \rightarrow +1} \Sigma_{\bullet}(\tau) = \Sigma_{\rm I(\bullet)}$.
	The stability around each fixed point is found again using the standard method to be given by
	\be 
	&& {\rm Eigen}(J^{\rm A})(\tau) = \left(- \frac{( 2 \sqrt{5} - 1)\tau}{3 \sqrt{5}}, - \frac{8 \tau}{3 \sqrt{5}} \right),\label{eq:eigenJA1} \\
	&& {\rm Eigen}(J^{\rm B})(\tau) = \left( - \frac{(2 \sqrt{5} + 1)\tau}{3 \sqrt{5}},  \frac{8 \tau}{3 \sqrt{5}} \right), \label{eq:eigenJB1}\\
	&& {\rm Eigen} ( J^{\rm C}) (\tau) =  \left( - \frac{( \sqrt{821}+21) \tau}{15}\label{eq:eigenJC1}
	,  \frac{( \sqrt{821} - 21 ) \tau}{15}  \right) .
	\ee
	Also, it is obvious from these formulas that we can, for example, read the instability index for each individual fixed point in the UV by calculating those in the IR and using the symmetry relation 
	\be
	{\rm Ind}[J^{U(\bullet)}] = 2 - {\rm Ind}[J^{I(\bullet)}],
	\ee
	across the boundary of past and future domains, i.e., $\tau=0$. We find
	\be
	&&   {\rm Ind}[J^A](\tau) =
	\begin{cases}
		2 & \mbox{for } \tau < 0 \\
		0 & \mbox{for } \tau > 0
	\end{cases}, \\
	&&   {\rm Ind}[J^B](\tau) =
	\begin{cases}
		1 & \mbox{for } \tau < 0 \\
		1 & \mbox{for } \tau > 0
	\end{cases}, \\
	&&   {\rm Ind}[J^C](\tau) =
	\begin{cases}
		1 & \mbox{for } \tau < 0 \\
		1 & \mbox{for } \tau > 0
	\end{cases}.
	\ee
	Fig. \ref{fig:flow_IS1} shows the flow structure on the $\tau$-fiber ${\cal M}_{\tau}$.
	
	Let us return to Eq.~(\ref{eq:IS_tau_3d}) for $\theta$, the flow defined on base space.
	As was mentioned above, the proportionality of eigenvalues in Eqs.~\eqref{eq:eigenJA1}-\eqref{eq:eigenJC1} with $\tau$ indicates
	that $\Sigma_{\rm A,B,C}(\tau)$ at $\tau=0$ change stability, which suggests that the Hartman-Grobman theorem~\cite{hartman,palmer1973generalization}~\footnote{This theorem, originally established for autonomous systems, states that one can use the linearization of the original dynamical system around a hyperbolic fixed point to analyze its behavior. A fixed point is said to be hyperbolic if none of the eigenvalues of Jacobian matrix is zero at that fixed point.} does not apply at this point where the kernel of the Jacobian matrix is nontrivial. Despite the fact that this, roughly speaking, tells us that the linearization of the dynamical system at $\tau=0$ will fail to capture the flow structure both qualitatively and quantitatively, nonetheless the flow still exists at this point in the extended configuration space simply because the dynamical system is regular there. Hence, there should exist a bounded invariant manifold ${\cal S} \subset {\cal M}$ for the flow such that
	\be
	\label{eq:invspace}
	{\cal S} := \{ p \in {\cal M} \ |\ && \lim_{\lambda \rightarrow -\infty} (\theta_{\lambda}(\tau), \phi(\lambda, p)) \in \{ -1 \} \times \Sigma_{\rm U} \ \mbox{ and } \nl
	&& 
	\  \lim_{\lambda \rightarrow +\infty} (\theta_{\lambda}(\tau),\phi(\lambda, p)) \in  \{ +1\} \times  \Sigma_{\rm I} \}.\label{eq:limSigma1}
	\ee
	where 
	\be
	&& \Sigma_{\rm U} := \{ \Sigma_{\rm U_{(A)}}, \Sigma_{\rm U_{(B)}} ,  \Sigma_{\rm U_{(C)}} \},  \qquad \Sigma_{\rm I} := \{ \Sigma_{\rm I_{(A)}} ,  \Sigma_{\rm I_{(B)}} ,  \Sigma_{\rm I_{(C)}} \}, \label{eq:Set_Sig}
	\ee
	denote the fixed point in the UV $(\rho\rightarrow -\infty)$ and IR $(\rho\rightarrow \infty)$, where $\Sigma_{\rm U,I(\bullet)}$ are given in Eq. (\ref{eq:crit_point_IS}).
	
	To investigate the relation between the invariant manifold and the associated fixed points on a more microscopic level, we define the projection map $\pi_{\cal S} : {\cal S} \rightarrow B$ from ${\cal S}$ onto the base space, and the $\tau$-fiber ${\cal S}_\tau$ is defined by $\pi_{\cal S}^{-1}(\tau)=: \{ \tau\} \times {\cal S}_{\tau}$ for $\tau \in B$ \footnote{In simpler terms, ${\cal S}_\tau$ is the evolving spacelike hypersurface of the invariant manifold $\cal S$, which is nothing but the {\it material hypersurface} of the flow described by the IS theory. This follows the language of fluid dynamics, in particular the theory of Lagrangian coherent structures (LCSs) which are
		codimension-one invariant manifolds in the extended configuration space of the time-aperiodic and finite-time nonautonomous systems \cite{haller2015lagrangian}.}. We observe that since $\Sigma_{\rm U,I_{(A)}}, \Sigma_{\rm U,I_{(B)}}, \Sigma_{\rm U,I_{(C)}}$ are isolated on ${\cal M}_{\tau=\mp 1}$, one can decompose the invariant manifold ${\cal S}$ into disconnected invariant submanifolds associated with the fixed points $\Sigma_{\rm U,I(\bullet)}$ and classify them.
	Formally, this classification might be carried out by introducing a structure group acting on the $\tau$-fiber ${\cal S}_{\tau}$.
	The procedure is explained as follows. Suppose that $U_{\alpha}$ and $U_{\beta}$ are open sets on $B$ with $U_{\alpha} \cap U_{\beta} \ne \emptyset$ and define a section as $\varphi_i : U_i \rightarrow \pi^{-1}_{\cal S}(U_i) = U_i \times {\cal S}_{U_i}$ for $i=\alpha,\beta$, where `$=$' stems from the fact that the fiber bundle is globally trivial.
	By constructing the composite mapping on $(U_{\alpha} \cap U_{\beta}) \times {\cal S}_{U_{\alpha} \cap U_{\beta}}$ as $\varphi_{\alpha} \circ \varphi_{\beta}^{-1}(\tau,{\bf X}) = (\tau,g_{\alpha \beta}(\tau){\bf X})$ for $\tau \in U_{\alpha} \cap U_{\beta}$, one can introduce a structure group over ${\cal S}$ as ${\rm Diff}({\cal S}) \subset {\rm Diff}({\mathbb R}^2)$ where
	\be
	{\rm Diff}({\cal S}) := \{ \, g_{\alpha \beta}(\tau) \in {\rm Diff(\mathbb R^2)}  \ | \  g_{\alpha \beta}(\tau) {\bf X} \in {\cal S}_{\tau}  \ \mbox{for all }  {\bf X} \in {\cal S}_{\tau} \, \} .
	\ee
	The structure group ${\rm Diff}({\cal S})$ can give away all there is to know about the disconnected submanifolds on the invariant manifold of the flows $S$, the number of which can be calculated from the topological properties of the quotient space ${\cal S}/{\rm Diff}({\cal S})$.
	According to the definition of ${\cal S}$ in Eq.~(\ref{eq:limSigma1}), overall there might exist at most nine possible disconnected invariant submanifolds such that
	\be
	{\cal S}_{ab} := \{ \,  S^{\prime} \subset {\cal S} \ | \  \lim_{\tau \rightarrow - 1} {\cal S}^{\prime}_{\tau} = \Sigma_{{\rm U}(a)}  \mbox{ and } \lim_{\tau \rightarrow + 1} {\cal S}^{\prime}_{\tau} = \Sigma_{{\rm I}(b)} \, \},  \label{eq:Sab}
	\ee
	where $a,b = {\rm  A,B,C}$ and $\Sigma_{\rm U,I(\bullet)}$ are given by Eq.~(\ref{eq:crit_point_IS}). Since the structure of fiber space changes at $\tau=0$, we decompose the invariant manifold ${\cal S}$ into the past and future domains in the following way:
	\be
	&&  {\cal S}_{a-} := \{ \,  S^{\prime} \subset \bigcup_{-1 \le \tau \le 0} \{ \tau \} \times {\cal S}_{\tau} \ | \  \lim_{\tau \rightarrow - 1} {\cal S}^{\prime}_{\tau} = \Sigma_{{\rm U}(a)}  \, \},   \\
	&&  {\cal S}_{a+} := \{ \,  S^{\prime} \subset \bigcup_{0 \le \tau \le +1} \{ \tau \} \times {\cal S}_{\tau} \ | \  \lim_{\tau \rightarrow + 1} {\cal S}^{\prime}_{\tau} = \Sigma_{{\rm I}(a)}  \, \}.
	\ee
	
	We consider the classification of the invariant manifold. This procedure essentially requites determining a partially ordered set -- so-called {\it poset} -- by introducing a partial order denoted by $<$ into an arbitrary set of fixed points. This procedure exhibits all the possible invariant manifolds by figuring out how the flow).
	
	In order to understand the flow structure of the Gubser flow, we would like to follow two crucial steps that will be fundamental to 
	the discussions of the existence of complete UV/IR flows and attractors to come:
	\begin{enumerate}
		\item
		We first give a basic mathematical definition for a pullback/forward attractor or repeller on the past and future domains. 
		A pullback/forward attractor or repeller constitutes a Morse set either on the past or future domain, which explains the (compact) invariant submanifolds of a nonautonomous flow.
		\item
		By gluing flows on the past and future domains using the notion of \textit{connecting sets} introduced in this paper, we establish the existence of flows on the invariant submanifolds satisfying Eq.~(\ref{eq:Sab}).
	\end{enumerate}
	
	Before doing this analysis, for later purposes we split the nonautonomous set into two subspaces since the flow in each space never intersects with the other. Thus we define
	\be
	&& {\cal M}^{T\ge 0} := \{ p \in {\cal M} \, | \, 0 \le T < +\infty\, \}, \\
	&& {\cal M}^{T \le 0} := \{ p \in {\cal M} \, | \, -\infty < T \le 0 \, \}.
	\ee
	One can see that the subspace with $T=0$ is closed under the flow which means that there exists no flow crossing the subspace.
	Thus, ${\cal M}^{T \ge 0}$ and ${\cal M}^{T \le 0}$ are also closed under the flows independently and one needs to look at only the one side. For that reason, we divide the total space into ${\cal M}^{T \lesseqgtr 0}$ as there is no flow across the boundary $T=0$; furthermore we separate them by their past / future domain attributed to the stability change near $\tau=0$. From now on, we would concentrate on the flow structure on ${\cal M}^{T \ge 0}$, but the structure of ${\cal M}^{T \le 0}$ can be obtained in a similar way described below. Apart from the division of the total space where analysis can be performed distinctively, it's also notable that in each subspace, the local \textit{forward/pullback} attractor needs to be distinguished for the flow in a nonautonomous system is characterized by both of its end points when it starts and finishes (see the following definition).
	
	Since a nonautonomous dynamical system does not respect time translation symmetry, for any family of solutions the limit $\tau\to\infty$ while $\tau_0$ is kept fixed could be different than the limit $\tau_0\to -\infty$ with $\tau$ kept fixed. Therefore, it becomes important to give a general definition for a local attractor and repeller \cite{Kloeden:2011,caraballo2017applied}, which creates a major difference from the well-studied autonomous dynamical systems (such as RG flow equations in field theories), where the space of all solutions itself does not change over time (or over energy scales). 
	\begin{kotak}
		\begin{definition} \label{def:att1}
			{\rm
				(Local pullback and forward attractor(repeller))\\
				Let $(\theta,\phi)$ be an invertible skew product flow.
				A non-empty compact and invariant nonautonomous set ${\cal A}=\bigcup_{\tau \in B} \{\tau \} \times A_{\tau} \subset {\cal M}$, i.e, $(\theta_{\lambda}(\tau),\phi(\lambda,{\cal A}))={\cal A}$ for all $\lambda  \in {\mathbb T}$, is called
				\begin{enumerate}
					\item a {\it local pullback attractor(repeller)} of $(\theta,\phi)$ if there exists $\epsilon \in {\mathbb R}^{+}$ such that
					\be
					\lim_{\lambda \rightarrow +\infty(-\infty)} {\rm dist}_{F}(\phi(\lambda,\theta_{-\lambda}(\tau),D_{\epsilon}(A_{\theta_{-\lambda}(\tau) }) ),A_{\tau}) = 0
					\ee
					holds for all non-positive(negative) $\tau \in B$;
					\item a {\it local forward attractor(repeller)} if there exists $\epsilon \in {\mathbb R}^{+}$ such that
					\be
					\lim_{\lambda \rightarrow +\infty(-\infty)} {\rm dist}_{F}(\phi(\lambda,\tau,D_{\epsilon}(A_{\tau})),A_{\theta_\lambda(\tau)}) = 0
					\ee
					holds for all non-negative(positive) $\tau \in B$,
				\end{enumerate}
				where $D_{\epsilon}({A}_{\tau})$ denotes an open disk of radius $\epsilon$ on the fiber space $F$ and ${\rm dist}_F$ is 
				defined as ${\rm dist}_F(A,B):=\sup_{a \in A} \inf_{b \in B}|a-b|$ for $A,B \subset F$ (Hausdorff semi-distance).
			}
		\end{definition}
	\end{kotak}
	In what follows, we take an Euclid metric to define a distance on the fiber space.
	The pullback attractor ${\cal R}$ and forward repeller ${\cal R}^*$ constitute an attractor-repeller(AR) pair $({\cal R},{\cal R}^*)$ which gives a Morse decomposition on the past domain \cite{Kloeden:2011}.
	In the same sense, the pullback repeller ${\cal A}$ and forward attractor ${\cal A}^*$ gives a pair $({\cal A},{\cal A}^*)$ providing a Morse decomposition on the future domain. 
	If one can find a pullback attractor (or repeller), it automatically means that there is a flow solution for which we can fix the current time and go back to history by sending the initial time $\tau_0\rightarrow-1$ or equivalently $\rho_0\rightarrow-\infty$. As an example, we plot in the Fig. \ref{fig:flow_IS1} different flow solutions. In there we observe that $\Sigma_{\rm U(A)}$ in the past domain can be intuitively identified as a local forward repeller $\mathcal R_{\rm A}$ while $\Sigma_{\rm U(B)}$ is associated with a local pullback attractor $\mathcal R_{\rm B}$.
	
	In the case of ${\cal M}^{T \ge 0}$, one can immediately find from Fig.~\ref{fig:flow_IS1} that there exists a flow $\Sigma_{\rm U(A)} \rightarrow \Sigma_{\rm U(B)}$, $\Sigma_{\rm U(A)} \rightarrow \Sigma_{\rm U(C)}$, and $\Sigma_{\rm I(B)} \rightarrow \Sigma_{\rm I(A)}$ on the fiber ${\cal M}^{T \ge 0}_{\mp 1}$. 
	Therefore, the AR pairs are
	\be
	\mbox{Past domain}&:& ({\cal R},{\cal R}^*) \, = \, ({\cal R}_{\rm C}, {\cal R}_{\rm A}), \, ({\cal R}_{\rm B}, {\cal R}_{\rm A}), \\
	\mbox{Future domain}&:& ({\cal A},{\cal A}^*) \, = \, ({\cal A}_{\rm B}, {\cal A}_{\rm A}),
	\ee
	where ${\cal R}_{a} \subseteq {\cal S}_{a-}$ is always identified in the past domain and ${\cal A}_{a} \subseteq {\cal S}_{a+}$ in the future domain satisfying Def.~\ref{def:att1}.
	These pairs naturally define the partial ordering among the fixed points,
	${\cal R}_{\rm C} <_- {\cal R}_{\rm A}, \, {\cal R}_{\rm B}<_-  {\cal R}_{\rm A}, {\cal A}_{\rm A} <_+ {\cal A}_{\rm B}$,
	where $<_{\mp}$ denotes the ordering on the past and future domains.
	
	To examine the existence of an invariant manifold, one needs to analyze the intersection of past / future domain where the flow coming from local AR pair become confluent. It is next required to glue the flow structure on the past and future domains. To this end, we define the partial order for invariant submanifolds on the past and future domains as follows
	\be
	{\cal S}_{a-, \tau = 0} \cap {\cal S}_{b+,\tau = 0} \ne \emptyset \ &\Rightarrow& \ {\cal A}_a < {\cal R}_b. 
	\label{eq:orde_PF}
	\ee
	Normally, computing the intersection in Eq.~(\ref{eq:orde_PF}) begs to find the local attractors and repellers, but instead we take a different approach in this study due to the difficulty of solving the ODEs generally. But before divulging the technical details, at least we can trivially verify that the existence of constant solutions $(T,\bar{\pi})=(0,\pm1/\sqrt{5})$ does indeed yield ${\cal A}_{\rm A} < {\cal R}_{\rm A}$ and ${\cal A}_{\rm B} < {\cal R}_{\rm B}$. 
	
\begin{figure}[htpb]
		\centering
		\begin{subfigure}[b]{0.45\textwidth}
			\centering
			\includegraphics[width=\textwidth]{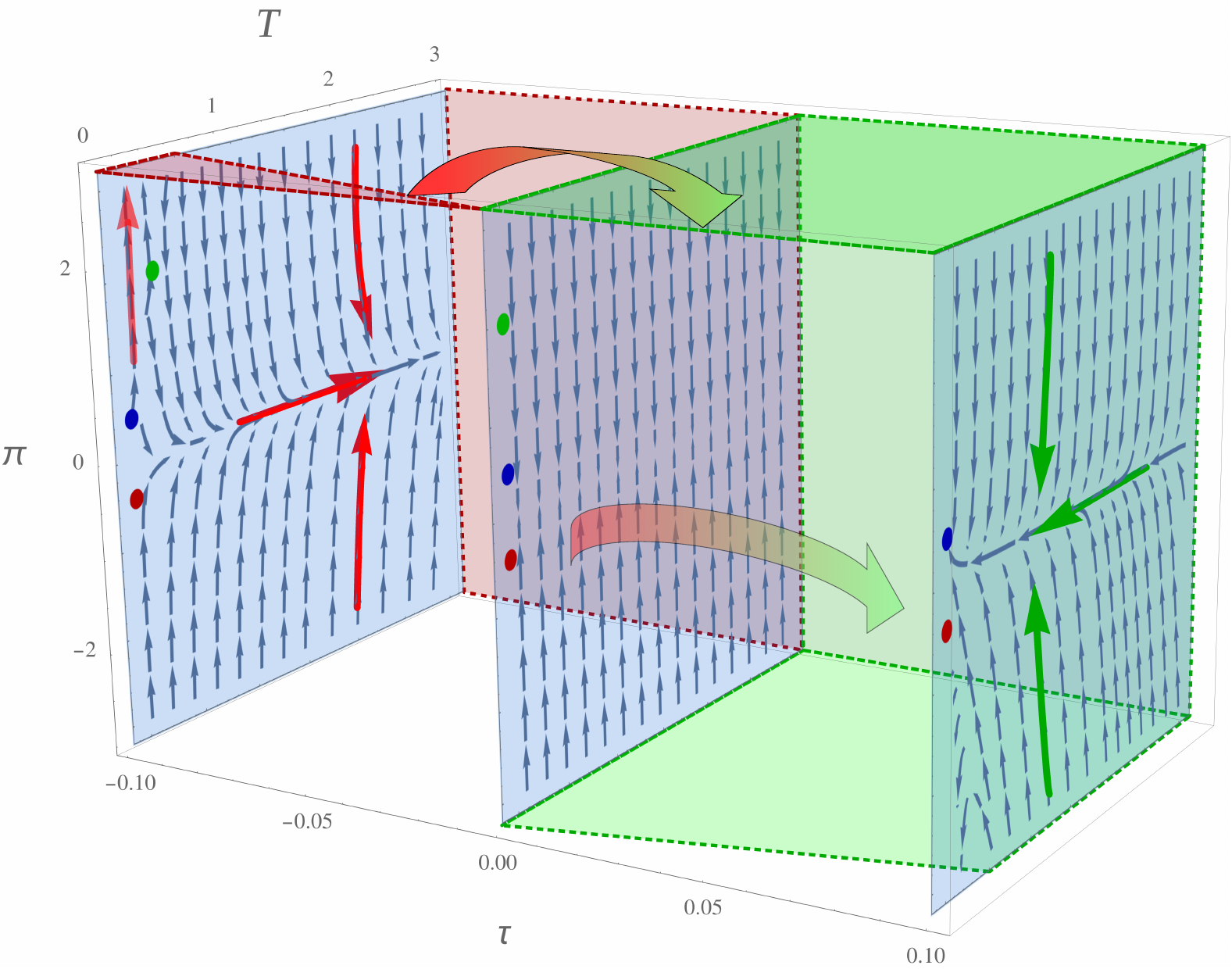} 
			\caption{${\mathcal M}^{T\geq 0}_{\pm \delta}$ and its connecting set near $\tau=0$}
			\label{fig:connectingTg0}
		\end{subfigure}
		\hfill
		\begin{subfigure}[b]{0.45\textwidth}
			\centering
			\includegraphics[width=\textwidth]{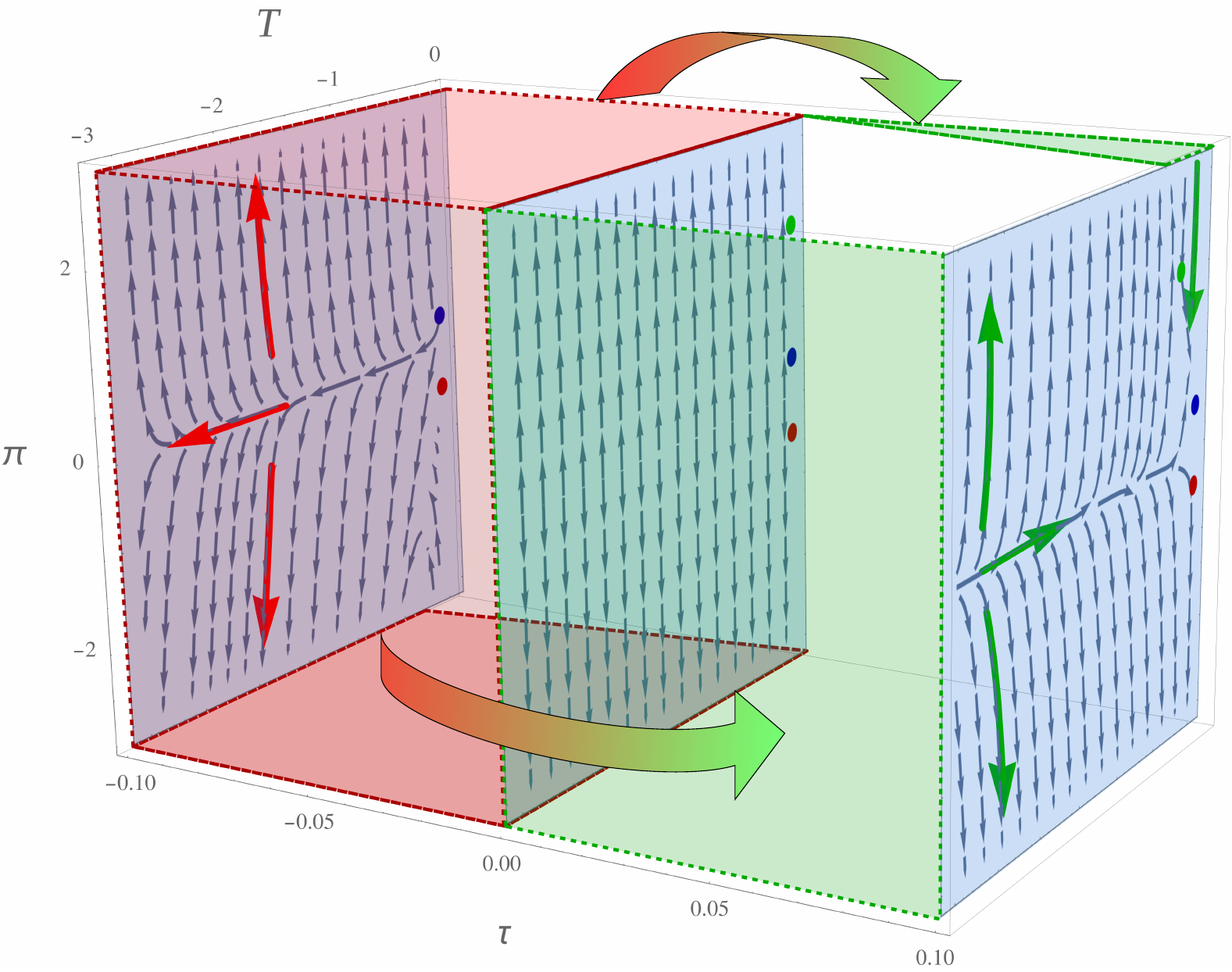} 
			\caption{${\mathcal M}^{T\leq 0}_{\pm \delta}$ and its connecting set near $\tau=0$}
			\label{fig:connectingTl0}
		\end{subfigure}
		\caption{Connecting sets of the fiber structures around $\tau=0$. Typical fiber structure $M_{\pm\delta}$ is shown in blue at $\tau=\{-0.1,0,0.1\}$ for the past and future domain, where the colored circles show the evolution of fixed point $\Sigma_A$ (blue), $\Sigma_B$ (red), and $\Sigma_C$ (green) on these fibers. The connecting set of each separate space ${\mathcal M}^{T\gtrless 0}$ is portrayed in red for $N_-$ and in green for $N_+$ where arrows indicate the directions of flow lines. The nontrivial poset for $M_{\pm \delta}^{T \ge 0}$ is determined as follows: We would take $\delta = 0.1$ and $\epsilon = 3$ for the connected subspace  $B_{\pm \delta}$. By looking at the flow directions around $\pd B_{\pm 0.1}$, the connecting set is approximately obtained as $N_{+} = ([0.5,3], -3) \cup (3,[-3,3]) \cup ([0,3], 3)$, $N_{-} =  (3, [-3,3]) \cup ([0,0.5],3)$, and thus $\tilde{N} = (3,[-3,3]) \cup ([0.5,3],3)$. On the one hand the in-coming flows intersecting with $\tilde{N}$ converge only to $\{ \Sigma_A \}$  on ${\cal M}^{T \ge 0}_{+0.1}$, on the other hand the out-going flows intersecting with $(3,[-3,3])$ and $([0.5,3],3)$ emerge from $\{ \Sigma_A,\Sigma_B \}$ and $\{ \Sigma_A,\Sigma_C \}$, respectively, on ${\cal M}^{T \ge 0}_{-0.1}$. Since the topology of $\pd B_{\pm \delta}$ does not change for $\delta>0.1$, the resulting poset is ${\cal A}_{\rm A} < {\cal R}_a$ for all $a \in \{\rm A,B,C \}$.
                  }
		\label{fig:connectingsets}
	\end{figure}

	In order to consider the other possibilities, we define a compact and connected subspace $B_{\pm \delta}$ on ${\cal M}^{T \ge 0}_{\pm \delta}$ with $0 < \delta \ll 1$ as\footnote{This range $\epsilon$ can be appropriately changed while keeping the topology of a closed disk.} 
	\be
	B_{\pm \delta} =  [0,\epsilon] \times [-\epsilon/2, \epsilon/2]  \subset F.
	\ee
	This region in the vicinity of hypersurface $\tau=0$ provides a layer where the past / future domain can be properly glued. To incorporate the flow reaching as far as the point $\Sigma_{\mathrm U(C)}$, we choose $\epsilon$ to be large enough such that $\Sigma_{\rm U(C)} \in {\rm Int}(B_{-\delta})$.
	Then, we introduce a \textit{connecting set} $\tilde{N}$ by
	\be
	&& \tilde{N} := N_{-} \cap N_{+} \subset F, \\
	&& N_{-} := \{ \, x \in  \pd B_{-\delta} \, | \, \mbox{transversally intersecting with out-going flows} \, \} ,  \\
	&& N_{+} := \{ \, x \in  \pd B_{+\delta} \, | \, \mbox{transversally intersecting with in-coming flows} \, \},
	\ee
	and $\tilde{N}$ lies on both ${\cal M}^{T \ge 0}_{\mp \delta}$.
	If there exists a flow connecting possibly any pair of $\Sigma_{\rm a}(-\delta)$ for $a={\rm A,B,C}$ and $\tilde{N}$ on ${\cal M}^{T \ge 0}_{-\delta}$, it can be deduced that $\tilde{N} < {\cal R}_{a}$.
	We should notice that any flow for $-1 < \tau \le -\delta$ starting at a $\Sigma_{\rm U(a)}$ can be continuously deformed into $\Sigma_{\rm a}(-\delta)$ due to the fact that the topology of $N_-$ does not change for any $\tau<0$, and therefore there cannot exist any singularities on the $\tau$-fiber.
	In a similar way, if there exists a flow which connects $\Sigma_{\rm a}(+\delta)$ and $\tilde{N}$ on ${\cal M}^{T \ge 0}_{+\delta}$, it can immediately lead to ${\cal A}_a < \tilde{N}$.
	Although the connecting set $\tilde{N}$ in ${\cal M}^{T \ge 0}_{- \delta}$ connects all $\Sigma_{\rm A,B,C}(-\delta)$ through flows at $\tau = -\delta$, only $\Sigma_{\rm A}(+\delta)$ connects to $\tilde{N}$ at $\tau = +\delta$. In Fig.\ref{fig:connectingsets} we illustrate how to define the connecting sets to obtain the posets.

	Following the discussion above, the nontrivial posets are then given by ${\cal A}_{\rm A} < {\cal R}_{\rm B}$ and ${\cal A}_{\rm A} < {\cal R}_{\rm C}$. Hence, we obtain the flow diagram on ${\cal M}^{T \ge 0}$ as
	\be
	\begin{tikzcd}
		{\cal R}_{\rm B} \arrow[rd]  \arrow[dd] &    {\cal R}_{\rm A} \arrow[l,dashed] \arrow[r,dashed] \arrow[d] \arrow[ddr] & {\cal R}_{\rm C} \arrow[ld]  \\
		& \tilde{N} \arrow [rd] & \\
		{\cal A}_{\rm B} \arrow [rr,dashed] &  & {\cal A}_{\rm A}  
	\end{tikzcd}
	\ \Rightarrow \
	\begin{tikzcd}
		{\cal R}_{\rm B} \arrow[rrdd]  \arrow[dd] &    {\cal R}_{\rm A} \arrow[l,dashed] \arrow[r,dashed]  \arrow[ddr] & {\cal R}_{\rm C} \arrow[dd]  \\
		& & \\
		{\cal A}_{\rm B} \arrow [rr,dashed] &  & {\cal A}_{\rm A}  
	\end{tikzcd} \label{eq:flow_Tg0}
	\ee
	where the dashed and solid arrows denote $<_{\mp}$ and $<$, respectively.
	
	In a similar way, the flow diagram on ${\cal M}^{T \le 0}$ is given by
	\footnote{
		For ${\cal M}^{T \le 0}$, we define the compact and connected subspace $B_{\pm \delta}$ by
		\be
		B_{\pm \delta} =  [-\epsilon,0] \times [-\epsilon/2, \epsilon/2]  \subset F.
		\ee
		and choose $\epsilon$ large enough such that $\Sigma_{\rm U(C)}(+\delta) \in {\rm Int}(B_{+\delta})$.
	}
	\be
	\begin{tikzcd}
		{\cal R}_{\rm B}  \arrow[dd] &  & {\cal R}_{\rm A}  \arrow[ll,dashed] \arrow[ld] \arrow[ldd] \\
		& \tilde{N} \arrow [rd] \arrow[d] \arrow[ld] & \\
		{\cal A}_{\rm B} \arrow[r,dashed] &    {\cal A}_{\rm A}    & {\cal A}_{\rm C} \arrow[l,dashed] 
	\end{tikzcd}
	\ \Rightarrow \
	\begin{tikzcd}
		{\cal R}_{\rm B}   \arrow[dd]   &  & {\cal R}_{\rm A} \arrow[dd] \arrow[lldd] \arrow[ldd]  \arrow[ll,dashed] \\
		& & \\
		{\cal A}_{\rm B} \arrow[r,dashed] & {\cal A}_{\rm A}  & {\cal A}_{\rm C}  \arrow[l,dashed]
	\end{tikzcd} \label{eq:flow_Tl0}
	\ee
	Finally, the invariant manifold ${\cal S}$ can be decomposed into six submanifolds as
	\be
	&& {\cal S} = {\cal S}_{\rm AA} \cup {\cal S}_{\rm AB} \cup {\cal S}_{\rm AC} \cup {\cal S}_{\rm BA} \cup {\cal S}_{\rm BB} \cup  {\cal S}_{\rm CA},\\
	&& {\cal S}_{\rm BC} = {\cal S}_{\rm CB} = {\cal S}_{\rm CC} = \emptyset.
	\ee
	Combining Eqs.~(\ref{eq:flow_Tg0}) and (\ref{eq:flow_Tl0}), the global picture of the flow structure is found to be
	\be
	\begin{tikzcd}
		{\cal R}_{\rm B}   \arrow[dd]  \arrow[ddr]  &   {\cal R}_{\rm A} \arrow[dd]  \arrow[ldd]  \arrow[l,dashed] \arrow[r,dashed] \arrow[rdd] & {\cal R}_{\rm C}   \arrow[ldd]  \\
		& & \\
		{\cal A}_{\rm B} \arrow[r,dashed] & {\cal A}_{\rm A}  & {\cal A}_{\rm C}  \arrow[l,dashed]
	\end{tikzcd} \label{eq:flow_full}
	\ee
For the purposes of this paper, the diagrams shown in \eqref{eq:flow_Tg0},~\eqref{eq:flow_Tl0} and-\eqref{eq:flow_full} reveal three important facts about the Gubser flow as follows. (1) They show which UV and IR fixed points are connected by a flow solution; (2) They classify these fixed points in terms of their local stability; (3) Finally, they determine the global structure of the space of solutions and whether there is an attractor/repeller in the past or future of the time-dependent system. Although these conclusions are specific to the present case, the methods discussed to obtain them could be established for any dynamical system, even those in which bifurcations and phase transitions could possibly exist.

	One of the interesting questions that one might address is how do we find the invariant submanifold ${\cal S}_{ab}$ having a nonempty set? This might be a very nontrivial question at first sight, but the transseries constructed in Sec.~\ref{sec:trans_const} does in fact provide a topological interpretation of the flow structure, thus indirectly answering our question. This follows next.

	\subsection{Transseries construction} \label{sec:trans_const} 
	In this subsection, we will consider the transseries around the stable IR fixed point (non-thermal equilibrium) $\Sigma_{\rm I(A)}$, but the transseries around $\Sigma_{\rm U(A)}$ (free streaming point) can be constructed with a similar strategy. We refer to Ref.~\cite{Costin:2008} for technical details.
	
	Let us identify first the appropriate expansion parameter. Since the radius of convergence for the expansion rate of the Gubser flow $D_\mu u^\mu=\tanh\rho$ is infinity, an asymptotic power expansion in terms of $1/\rho$ would not work. Furthermore, as we have shown in the previous section, the leading-order behavior of the asymptotic solutions for the IS theory is determined by Eqs.~\eqref{eq:IS_T_sle} and \eqref{eq:IS_pi_sle}. Thus, the correct expansion parameter which captures the IR limit is obtained by $z \rightarrow +\infty$. Rewriting
	the ODEs of (\ref{eq:IS_T}) and (\ref{eq:IS_pi}) in $z$ variable, we get
	\begin{subequations}
		\label{eq:IS_z}
		\begin{align}
		&\frac{d T}{d z} = \frac{T}{6z } (\bar{\pi} - 2 ) \frac{1-z^{-1}}{1+z^{-1}}, \label{eq:IS_T_z}  \\
		&\frac{d \bar{\pi}}{d z} = - \frac{2}{3z} \left( \bar{\pi}^2 - \frac{1}{5} \right) \frac{1-z^{-1}}{1+z^{-1}} - \frac{T \bar{\pi}}{2cz}. \label{eq:IS_pi_z} 
		\end{align}
	\end{subequations}
	For the sake of simplicity, we shift the variables $T \rightarrow T + T^c$ and $\bar{\pi} \rightarrow \bar{\pi} + \bar{\pi}^c$ where the constants $\bar{\pi}^c,T^c$ stand for the coordinates of the fixed point $\Sigma_{\rm I(A)}$ given by Eq.~(\ref{eq:crit_point_IS}). To our benefit, this transformation has brought the fixed point to the origin of configuration space ($T \rightarrow 0, \bar{\pi} \rightarrow 0$ as $z \rightarrow +\infty$). As a result, the modified version of the ODEs~\eqref{eq:IS_z} is 
	\be
	&& (z+1) \frac{d{\bf X}}{d z}
	= {\frak B} {\bf X} + {\frak C}({\bf X}) {\bf X} + z^{-1} {\frak D}({\bf X}) {\bf X}, \label{eq:diff_IS_X}
	\ee
	where ${\bf X} = (T,\bar{\pi})^{\top}$, and the matrices are given by
	\be
	&& {\frak B} =
	\begin{pmatrix}
		\frac{1}{6} \left( \frac{1}{\sqrt{5}} -2 \right)  & 0 \\
		- \frac{1}{2 \sqrt{5} c} & -\frac{4}{3 \sqrt{5}}
	\end{pmatrix}, \\
	&& {\frak C}({\bf X}) = 
	\begin{pmatrix}
		\frac{\bar{\pi}}{6}   & 0 \\ 
		0  &  - \frac{2 \bar{\pi}}{3}  -\frac{T}{2c} 
	\end{pmatrix}, \\
	&& {\frak D}({\bf X}) = 
	\begin{pmatrix}
		- \frac{1}{6} \left( \frac{1}{\sqrt{5}} -2 + \bar{\pi} \right)     & 0 \\ 
		-\frac{1}{2\sqrt{5}c} & \frac{4}{3 \sqrt{5}} + \frac{2 \bar{\pi}}{3}  -\frac{T}{2c} 
	\end{pmatrix}.
	\ee
	Next, we diagonalize ${\frak B}$ by way of multiplying the ODE in (\ref{eq:diff_IS_X}) by an invertible matrix $U$ from the left:
	\be
	&& (z+1) \frac{d \tilde{\bf X}}{d z}
	= \tilde{\frak B} \tilde{\bf X} + \tilde{\frak C}({\bf X}) \tilde{\bf X} + z^{-1} \tilde{\frak D}({\bf X}) \tilde{\bf X}, \label{eq:diff_Xtil} \\
	&& \tilde{\bf X} = U{\bf X}, \\ 
	&& \tilde{\frak B} = U {\frak B} U^{-1} = {\rm diag}(b_1,b_2) \, = \, {\rm diag} \left( -\frac{1}{3} + \frac{1}{6 \sqrt{5}},  - \frac{4}{3 \sqrt{5}} \right), \label{eq:IS_diagb}\\
	&& \tilde{\frak C}({\bf X}) = U {\frak C}({\bf X}) U^{-1} =
	\begin{pmatrix}
		\frac{\bar{\pi}}{6} & 0 \\
		\frac{5 \bar{\pi}}{6} + \frac{T}{2 c} & - \frac{2\bar{\pi}}{3} - \frac{T}{2 c}
	\end{pmatrix} , \\
	&& \tilde{\frak D}({\bf X}) = U {\frak D}({\bf X}) U^{-1} =
	\begin{pmatrix}
		\frac{1}{3}- \frac{1}{6 \sqrt{5}} - \frac{\bar{\pi}}{6} & 0 \\
		\frac{2}{3} - \frac{3}{\sqrt{5}}  - \frac{5 \bar{\pi}}{6} + \frac{T}{2 c} & \frac{4}{3 \sqrt{5} } + \frac{2\bar{\pi}}{3}  - \frac{T}{2 c}
	\end{pmatrix},
	\ee
	where the invertible matrix $U$ is given by
	\be
	&& U =
	\begin{pmatrix}
		1 & 0 \\
		1 &  \frac{(9 - 2 \sqrt{5}) c}{3}
	\end{pmatrix}, \qquad
	U^{-1} =
	\begin{pmatrix}
		1 & 0 \\
		-\frac{3}{(9 - 2 \sqrt{5})c}  &  \frac{3}{(9 - 2 \sqrt{5})c}
	\end{pmatrix}.
	\ee
	Notice that $T$ and $\bar{\pi}$ in $\tilde{\frak C}({\bf X})$ and $\tilde{\frak D}({\bf X})$ can be simply obtained using ${\bf X}=U^{-1} \tilde{\bf X}$.
	
	The transseries ansatz is constructed from the solution of the linearized ODEs (\ref{eq:IS_T_sle}) and (\ref{eq:IS_pi_sle}). These solutions can then be directly taken as transmonomials of the transseries, which essentially encode the information of nonperturbative contributions to the perturbative expansion in $1/z$.
	Hence, the transseries ansatz is given by
	\be
	&& \tilde{\bf X}(z) = \sum_{|{\bf m}|\ge 0}^\infty \sum_{k=0}^\infty \tilde{\bf X}_{k}^{({\bf m})} \Phi^{{\bf m}}_{k}, \qquad \mbox{as \ } z \rightarrow +\infty, \label{eq:X_trans} \\
	&& \Phi^{\bf m}_{k} := \bm{\sigma}^{\bf m} z^{{\bf m} \cdot \bm{\beta}-k}, \qquad \bm{\sigma}^{\bf m}:= \sigma_1^{n_1} \sigma_2^{n_2},
	\ee
	where ${\bf m} \in {\mathbb N}_0^2$, $\bm{\beta} \in {\mathbb R}^2$, and $\bm{\sigma} \in {\mathbb R}^2$ are the integration constants.
	The ansatz (\ref{eq:X_trans}) can be regarded as an element of a vector space ${\cal V}$ over ${\mathbb R}$ equipped with the basis $\Phi^{\bf m}_k (k \in {\mathbb Z})$.
	Since the basis $\Phi^{\bf m}_{k}$ satisfies the properties
	\be
	&& z^{-k^\prime} \Phi^{\bf m}_{k} = \Phi^{\bf m}_{k+k^{\prime}}, \\
	&& \Phi^{\bf m}_{k} \cdot \Phi^{{\bf m}^\prime}_{k^{\prime}} = \Phi^{{\bf m}+{\bf m}^{\prime}}_{k+k^{\prime}}, \\
	&& \frac{d \Phi^{\bf m}_{k}}{d z} = ({\bf m} \cdot \bm{\beta}-k)  \Phi^{\bf m}_{k+1},
	\ee
	the vector space ${\cal V}$ is said to form an ${\mathbb R}$-polynomial ring ${\mathbb R}[z^{-1},\sigma_1 z^{\beta_1},\sigma_2 z^{\beta_2}]$, that is closed under all the operations allowed in the ODE.
	Notice that $\tilde{\frak C}({\bf X})$ and $\tilde{\frak D}({\bf X})$ in (\ref{eq:diff_Xtil}) might be expanded in the same basis as
	\be
	\tilde{\frak C}({\bf X}) = \sum_{|{\bf m}|\ge 0}^\infty \sum_{k=0}^\infty \tilde{\frak C}_{k}^{({\bf m})} \Phi^{\bf m}_{k},  \qquad \tilde{\frak D}({\bf X}) = \sum_{|{\bf m}|\ge 0}^\infty \sum_{k=0}^\infty \tilde{\frak D}_{k}^{({\bf m})} \Phi^{\bf m}_{k}. \label{eq:CD_trans}
	\ee
	Substituting the ansatz (\ref{eq:X_trans}) and (\ref{eq:CD_trans}) in (\ref{eq:diff_Xtil}), we find the evolution equation for the transseries coefficients as
	\be
	&& ({\bf m}\cdot \bm{\beta}-k) \tilde{X}_{i,k}^{({\bf m})} + ({\bf m}\cdot \bm{\beta}-k+1) \tilde{X}_{i,k-1}^{({\bf m})} \nl
	&=& b_{i} \tilde{X}_{i,k}^{({\bf m})} \Phi^{\bf m}_{k} + \sum_{j=1}^{2} \sum_{|{\bf m}^{\prime}| \ge 0}^{\bf m} \left( \sum_{k^\prime=0}^{k} \tilde{\frak C}_{ij,k^\prime}^{({\bf m}^{\prime})} \tilde{X}_{j,k-k^\prime}^{({\bf m}-{\bf m}^{\prime})} + \sum_{k^\prime=0}^{k-1} \tilde{\frak D}_{ij,k^\prime}^{({\bf m}^{\prime})} \tilde{X}_{j,k-k^\prime-1}^{({\bf m}-{\bf m}^{\prime})} \right), \label{eq:IS_evo}
	\ee
	where $b_i$ is already written in Eq.~(\ref{eq:IS_diagb}).
	We next normalize the integration constants as $\tilde{X}_{1,0}^{((1,0))} = \tilde{X}_{2,0}^{((0,1))} = 1$, all of $\tilde{X}_{j,k}^{({\bf m})}$ and $\beta_i$ are determined \textit{unambiguously} due to the lack of any singularities on the Borel plane.
	We calculate from (\ref{eq:IS_evo}) the anomalous dimensions $\beta_i = b_i$ and $\tilde{\bf X}^{({\bf 0})}_k = {\bf 0}$ for any $k$.
	These put together the final pieces of the puzzle for the transseries solution $\tilde{\bf X}(z)$ of Eq.(\ref{eq:IS_evo}), from which the formal transseries solution to \eqref{eq:diff_IS_X} is found to be
	\be
	&& {\bf X}(z) = \sum_{|{\bf m}| > 0}^\infty \sum_{k=0}^\infty {\bf X}_{k}^{({\bf m})} \Phi^{\bf m}_{k} + {\bf X}^c, \label{eq:X_transsol} \\
	&&    {\bf X}_{k}^{({\bf m})}  = U^{-1} \tilde{\bf X}_{k}^{({\bf m})} .
	\ee
	Once again since ${\bf X}_{0}^{({\bf 0})} = 0$, the fact that the formal transseries does not include exponential transmonomials is sufficient to prove that we have a \textit{convergent series}.
	
	Now the formal transseries expanded around the UV fixed point $\Sigma_{\rm U(A)}$ given by Eq.(\ref{eq:crit_point_IS}) is calculated using the time coordinate $\tilde{z}=1/z$. The UV limit is attained by $\tilde{z} \rightarrow +\infty$, and the ODE is given by
	\begin{subequations}
		\label{eq:IS_z_UV}
		\begin{align}
		&\frac{d T}{d \tilde{z}} = \frac{T}{6\tilde{z} } (\bar{\pi} - 2 ) \frac{1-\tilde{z}^{-1}}{1+\tilde{z}^{-1}}, \label{eq:IS_T_z_UV}  \\
		&\frac{d \bar{\pi}}{d \tilde{z}} = - \frac{2}{3\tilde{z}} \left( \bar{\pi}^2 - \frac{1}{5} \right) \frac{1-\tilde{z}^{-1}}{1+\tilde{z}^{-1}} + \frac{T \bar{\pi}}{2c\tilde{z}}. \label{eq:IS_pi_z_UV}
		\end{align}
	\end{subequations}
	Note that the difference between Eqs.~\eqref{eq:IS_z} and Eqs.~\eqref{eq:IS_pi_z_UV} is that the sign of last term in the second line of the former. Therefore, given the formal transseries (\ref{eq:X_transsol}), it would possible to directly turn it into the one around the UV fixed point $\Sigma_{\rm U(A)}$ by flipping the sign of relaxation scale $c \rightarrow -c$. Indeed, the transformation $z \rightarrow \tilde{z}$ together with $c \rightarrow -c$ gives an isomorphism between the transseries, i.e. ${\mathbb R}[z^{-1},\sigma_1 z^{\beta_1},\sigma_2 z^{\beta_2}]\rightarrow{\mathbb R}[\tilde{z}^{-1},\sigma_1 \tilde{z}^{\beta_1},\sigma_2 \tilde{z}^{\beta_2}]$.
	
	\subsection{Double expansion of the transseries}
	
	The expansion parameter of the UV/IR transseries derived previously is determined by the asymptotic properties of the IS theory. However, it is instructive to know the transseries in Knudsen number defined as $w=\tanh \rho /T(\rho)$. In this case, we rewrite the evolution equation from the original ODEs in (\ref{eq:IS_T}) and (\ref{eq:IS_pi}) in terms of $w$. As shown in Ref.~\cite{Behtash:2017wqg}, this variable does not lead to the correct asymptotic behavior. However, it is possible to construct a formal transseries by following the procedure outlined below.
	
	Instead of trying to solve Eqs. (\ref{eq:IS_T}) and (\ref{eq:IS_pi}), one can first think of completely eliminating $T(\rho)$ from these ODEs by introducing the parameter $w=\tanh \rho /T(\rho)$, that gives rise to the dynamical system~\cite{Behtash:2017wqg}
	\be
	&& \frac{\pd {\cal A}(w,\rho)}{\pd w} = F_{\cal A} ({\cal A}(w,\rho),w,\rho), \label{eq:IS_w} \\
	&& F_{\cal A} ({\cal A}(w,\rho),w,\rho) :=  -\frac{  \frac{4}{3} \left( 3 {\cal A}(w,\rho) + 2  \right)^2 + \frac{3 {\cal A}(w,\rho)+2}{c w} - \frac{4}{15}}{ 3 w\left( \coth^2 \rho -1 -{\cal A}(w,\rho) \right)}, 
	\ee
	where ${\cal A}(w,\rho)$ is defined by
	\be
	&& {\cal A}(w,\rho) = \frac{d \log T(w,\rho)}{d \log(\cosh \rho)},\qquad \bar{\pi}(w,\rho) = 3 {\cal A}(w,\rho)+2.
	\ee
	
	Trying to construct the transseries solution for the nonlinear ODE (\ref{eq:IS_w}), we notice the appearance of both $\rho$ and $w$ 
	in \eqref{eq:IS_w} simultaneously. Although a transseries solution depending on two variables is available as asymptotic \textit{double expansion} in terms of two generators $1/z=e^{- 2 \rho}$ and $1/w$, we need to include a $\log$-type transmonomial as well.
	
	Let us suppose that the flows converge to the IR fixed point in the $w,z \rightarrow +\infty$ limit.
	Given the IR fixed point of Eq.~\eqref{eq:IS_w}, ${\cal A}^c=1/(3\sqrt{5})-2/3$, we can  linearize the ODE of (\ref{eq:IS_w}) at 
	this value using
	\be
	&& \frac{\pd {\cal A}(w,z)}{\pd w} = \left. \frac{\pd F_{\cal A}({\cal A},w,z)}{\pd {\cal A}} \right|_{{\cal A} \rightarrow {\cal A}^c} \delta {{\cal A}(w,z)},
	\ee
	in which 
	\be
	\label{eq:IS_fluc}
	\delta {\cal A}(w,z) &=& \sigma w^{8/(\sqrt{5}+1-3 \sqrt{5} \coth^{2} \rho) } \nl
	&\sim& \sigma w^{\beta} \left[ 1 +  \frac{480 \sqrt{5} \log w }{(-10 + \sqrt{5})^2 z}  \right. \nl
	&& \qquad \qquad \left. + \frac{4800 \left\{  \left( 10 - 39 \sqrt{5} \right) \log w +120 \left( \log w \right)^2 \right\}}{(-10 + \sqrt{5})^4 z^2} + \cdots   \right], 
	\ee
	where $\beta = -8(1+2 \sqrt{5})/19$ and $\coth \rho=(1+z^{-1})/(1-z^{-1})$. Eq.~\eqref{eq:IS_fluc} shows that the transseries of $\mathcal A(w,z)$ is an asymptotic expansion with leading-order term $w^\beta$ followed by corrections characterized by $1/w,\, \log w\,\text{ and } 1/z$. Here, the initial condition is decoded in the integration constant $\sigma$. This result indicates that the formal transseries solution of Eq.~\eqref{eq:IS_w} can be expressed in the form
	\be
	{\cal A}(w,z) &=& {\cal A}^c + \sum_{k=1}^{+\infty} \sum_{h=0}^{+\infty} {\cal A}^{(0)}_{(k,0,h)} w^{-k}  z^{-h}   + \sum_{n=1}^{+\infty} \sum_{k,h=0}^{+\infty} \left( \sigma w^{\beta}\right)^n {\cal A}^{(n)}_{(k,0,h)} w^{-k} z^{-h}   \nl
	&& + \sum_{n=1}^{+\infty} \sum_{k=0}^{+\infty} \sum_{s,h=1}^{+\infty}  \left( \sigma w^{\beta}\right)^n {\cal A}^{(n)}_{(k,s,h)} w^{-k} (\log w)^s z^{-h}, \qquad \quad \mbox{as \ } w,z \rightarrow +\infty, \nl \label{eq:IS_w_z}
	\ee
	where ${\cal A}_{(k,s,h)}^{(n)} \in {\mathbb R}$ are the expansion coefficient and $\sigma \in {\mathbb R}$.
	It is important to emphasize that the Borel transform of \eqref{eq:IS_w_z} even in the presence of $\log w$ does not have singularities on the Borel plane. Again, we want to stress that having no exponential corrections in the expansion is a direct indicator of 
	being perturbative in nature and therefore \eqref{eq:IS_w_z}  should be convergent.
	To check this out quickly, we take $z \rightarrow +\infty$ but keep $w$ finite, i.e. fix $h=0$ in the solution (\ref{eq:IS_w_z}).
	Then, the formal solution becomes
	\be
	{\cal A}(w,+\infty) &=& {\cal A}^c + \sum_{k=1}^{+\infty} {\cal A}^{(0)}_{(k,0,0)} w^{-k}   + \sum_{n=1}^{+\infty} \sum_{k=0}^{+\infty} \left( \sigma w^{\beta}\right)^n {\cal A}^{(n)}_{(k,0,0)} w^{-k} , \qquad \mbox{as \ } w \rightarrow +\infty. \nl \label{eq:IS_w_zinf}
	\ee
	One can readily find that this series is convergent. As far as Borel summability of ${\cal A}(w,+\infty)$ goes, a useful way is to take the inverse Laplace transform of the ODE (\ref{eq:IS_w}). Suppose that ${\cal A}(w,+\infty)$ is an asymptotic power series about the IR fixed point, i.e. ${\cal A}(w,+\infty)=\sum_{k=0}^{+\infty}{\cal A}^{(0)}_{(k,0,0)} w^{-k}$.
	After applying the shift $A(w,+\infty) \rightarrow A(w,+\infty)+ {\cal A}^c$, the inverse Laplace transformed ODE becomes
	\footnote{
		Notice that $(c_1 * c_2)(p) = O(p)$ for constant $c_1$ and $c_2$,
		because
		\be
		p^{t} * p^s = \int^{p}_{0} dq \, q^t (p-q)^s =  \frac{\Gamma(t+1)\Gamma(s+1)}{\Gamma(t+s+2)} p^{t+s+1}, \qquad \Re (t) > -1, \ \Re (s) > -1.
		\ee
	}
	\be
	&& -p Y(p) =
	\sum_{s=0}^{+\infty} \frac{(-15)^{s} }{\left(-10+\sqrt{5} \right)^{s+1}}
	\left[ \frac{1}{c}  \left\{ \sqrt{5} p  + 15 (p*Y)(p) \right\} \right. \nl
	&& \qquad \qquad   \left. +  60 (1 * Y^{*2})(p) +  8 \sqrt{5} (1 *  Y)(p)  \right](* Y(p))^s, \label{eq:IS_w_Borel}
	\ee
	where $p$ is the Borel parameter, and ${\cal B}$ stands for the Borel transform defined by
	\be
	&& Y(p) := {\cal B} {\cal A}(w,+\infty),  \\
	&& {\cal B} w^{k-1} = \frac{p^k}{\Gamma(k+1)}.
	\ee
	Also, $*$ denotes the convolution operation
	\be
	&& (G_1 * G_2) (p) := \int^{p}_{0} dq \, G_1(q) G_2 (p-q), \label{eq:convo} \\
	&& G_1^{*s}(p) := (\underbrace{G_1 * \cdots * G_1}_{s \ \mbox{times}})(p).
	\ee
	The original asymptotic series ${\cal A}(w,+\infty)$ can be reproduced by computing the Laplace integral of $Y(p)$, 
	\be
	&&\int_{0}^{\infty e^{i \theta} }dp \, e^{-p w} \, Y(p) \nl
	&=& -\sum_{s=0}^{+\infty} \int_{0}^{\infty e^{i\theta}}dp \, e^{-p w} \,
	\frac{(-15)^{s} }{\left(-10+\sqrt{5} \right)^{s+1}p} 
	\left[ \frac{1}{c}  \left\{ \sqrt{5} p  + 15 (p*Y)(p) \right\} \right. \nl
	&& \qquad \qquad   \left. +  60 (1 * Y^{*2})(p) +  8 \sqrt{5} (1 *  Y)(p)  \right](* Y(p))^s, \qquad \theta \in [0,2 \pi). \label{eq:laplase_A}
	\ee
	If $Y(p)$ is not Laplace transformable for some $\theta$, all the terms on the r.h.s. of Eq.~(\ref{eq:laplase_A}) must entail a singularity across the board, but here they do not for any arbitrary value of $\theta$.
	Therefore, $Y(p)$ is an entire function on the Borel plane and thus, the asymptotic series of ${\cal A}(w,+\infty)$ converges~\footnote{
		For example, if one considers
		\be
		\frac{d f(w)}{d w} = - \lambda f(w) - w^{-1} ( f(w) + 1 ) 
		, \qquad \lambda \in {\mathbb R}^+ , \label{eq:ex_f}
		\ee
		the inverse Laplace transformed equation is given by
		\be
		-p F(p) = - \lambda F(p) - (1 * F)(p) -1,  
		\ee
		where ${\cal B}f(w)=F(p)$.
		In this case, there exists a singularity at $p=\lambda$ because
		\be
		F(p) =  \frac{1 +  (1 * F)(p) 
		}{p-\lambda}.
		\ee
		Indeed, the Borel transformed asymptotic series of the ODE (\ref{eq:ex_f}) is Borel nonsummable. This explains why the asymptotic expansion series for the Bjorken flow is Borel nonsummable since the linearized differential equation for the function $\mathcal{A}$ is of the form~\eqref{eq:ex_f}~\cite{Behtash:2018moe,Behtash:2019txb}. 
	}.

	\section{Moment expansion method for the kinetic Gubser flow}
	\label{sec:mthofmoments}
	
	In this section, we generalize the ideas of nonautonomous dynamical systems to the case of the Boltzmann equation within the relaxation time approximation (RTA) for the Gubser flow. In what follows, the problem of solving the Boltzmann equation is basically mapped to the study of a dynamical system over the space of moments of the distribution function and de Sitter time. 
	
	The symmetry group of the Gubser flow, $SO(3)_q\times  SO(1,1)\times  {\mathbb Z}_2$, restricts the number of variables on which the distribution function depends~\cite{Denicol:2014xca,Denicol:2014tha}. It was demonstrated that the on-shell distribution function for this flow is given by $f(x^\mu,p_i)=f(\rho,p_\Omega^2,|p_\varsigma|)$~\cite{Denicol:2014xca,Denicol:2014tha}. Thus, the RTA Boltzmann equation is reduced to the following relaxation equation~\cite{Denicol:2014xca,Denicol:2014tha}
	\be
	\frac{\pd f(\rho,p_\Omega^2,|p_\varsigma|)}{\pd \rho} = -\frac{1}{\tau_R(\rho)} \left[ f(\rho,p_\Omega^2,|p_\varsigma|) - f_{\rm eq}(p_{\rho}/{T}(\rho)) \right], \label{eq:Boltzeq_rho}
	\ee
	in which $\tau_R(x):=\theta_{0}/{T}(x)$ is the relaxation time scale, which is taken to be a positive constant. In Eq.~\eqref{eq:Boltzeq_rho}, the angular momentum $p_\Omega^2$ and the energy $p_\rho$ are given by
	\begin{subequations}
		\begin{align}
		p_\Omega^2 &= p_\theta^2\,+\,\frac{p_\phi^2}{\sin^2\theta}\,,\\
		p_\rho&=\sqrt{\frac{p_\Omega^2}{\cosh^2\rho}+p_\varsigma^2}\,,
		\end{align}
	\end{subequations}
	in which $p_\varsigma$ is the longitudinal momentum in the boosted frame. We assume that the equilibrium distribution function has a functional form of the Maxwell-Boltzmann type, i.e. $f_{\rm eq}(x)=e^{-x}$. In statistical mechanics it is known that for systems invariant under a certain scaling, the zeros of the collisional kernel $\mathcal{C}[f] = 0$ admit time-independent solutions, often called steady states, which do not correspond to the global {\it thermal} equilibrium state for which $f(x,p) = f_{\rm eq}(p_{\mu} \beta^{\mu}(x))$. As we shall see, the stability analysis for the moments of the distribution function indicates the dynamical equilibrium state is not a thermal one; rather it is a non-thermal steady state. This is at odds with the equilibrium state for the Bjorken flow~\cite{Behtash:2019txb}. Hence, even though the equilibrium distribution function satisfies $\mathcal{C}[f]=0$,
	there might be another zero of the collisional kernel that defines a global equilibrium state~\cite{zakharov2012} \footnote{In the RTA approximation, temperature is an overall function in the collisional kernel. So even if $f-f_{\rm eq}$ does not vanish as time tends to infinity, the vanishing of temperature guarantees that we have a global non-thermal equilibrium for the Boltzmann equation.}.  
	
	We expand the distribution function in terms of orthogonal polynomials~\footnote{In our previous publications~\cite{Behtash:2018moe,Behtash:2019txb} a similar ansatz was introduced for the Bjorken flow. A reader familiar with representation theory would recognize that the generalized Laguerre polynomials are a representation of $SU(1,1) \cong SO(2,1)$ ~\cite{GROENEVELT2003329}, which is not a subgroup of the spacetime symmetry group of the Bjorken flow. Nonetheless, since $SO(1,1)\cong SO(2,1)/{\frak G}$ where ${\frak G}=SO(1,1)\times SO(2)$, there should be some symmetry breaking conditions in going to the momentum space that would reduce the $SO(2,1)$ down to $SO(1,1)$, which is in part fulfilled with constraints coming from the geometry of spacetime and conservation laws. It would then be more appropriate to say that the symmetries of $f$ are in general those of the phase space from the perspective of Hamiltonian dynamics. Yet, here the expansion in \eqref{eq:boltz_Gub} corresponds to the genuine representation theory of the momentum space symmetries whose generators commute with a Hamiltonian that should come from the Boltzmann equation. But due to the collisional kernel being 
		a nonlinear operator, we cannot get this Hamiltonian for the problem at hand. Thus, we are at a loss to say anything on the representation theoretical aspects of \eqref{eq:boltz_Gub} and would rather trust the fact that it {\it happens} to solve the Boltzmann equation.} 
	\be
	f(\rho,p_\Omega^2,p_\varsigma^2) = e^{-p_{\rho}/{T}(\rho)}\,\sum_{n,\ell=0}^{+\infty} c_{n \ell}(\rho) {\frak P}_{2 \ell}(\cos \theta_p) {\frak L}^{(3)}_{n}\left( \frac{p_{\rho}}{{T}(\rho)} \right)\,, \label{eq:boltz_Gub}
	\ee
	where ${\frak P}_\ell(x)$ and ${\frak L}^{(3)}_{n}(x)$ are the Legendre and generalized Laguerre polynomials, respectively. The variables in the polynomial basis of the expansion and its exponential weight factor are constrained by the symmetry of the flow. The angle variable $\theta_p$ in Eq.~\eqref{eq:boltz_Gub}  is determined by $\cos \theta_{p} = p_{\varsigma}/p_{\rho}$.  The coefficients $c_{nl}(\rho)$ can be projected out by the orthogonality of moments, and thus
	\be
	c_{n\ell}(\rho) = \frac{2 \pi^2 (4 \ell +1)}{T(\rho)^4} \frac{  \Gamma(n+1)}{  \Gamma(n+4)} \left\langle p^2_{\rho} \, {\frak P}_{2 \ell} (\cos \theta_p)  {\frak L}^{(3)}_{n} \left( \frac{p_{\rho}}{T(\rho)}\right)  \right\rangle_{f},
	\label{eq:cnl}
	\ee
	where $\langle {\cal O}(x,p) \rangle_{f} := \int_p \, {\cal O}(x,p) f(p)$ 
	in which we have defined
	\be
	\int_p := \int \frac{d^4 p}{(2 \pi)^4} \frac{1}{\sqrt{- \det g}}(2 \pi)\delta(p^2) 2 \theta(p_{\rho}) \,= \int \frac{d^3 {\bf p}}{(2 \pi)^3 p_{\rho} \cosh^2 \rho \sin \theta},
	\ee
	with $\theta(x)$ being the step function. Notice that for the Maxwell-Boltzmann type distribution, we have to set $c_{00}(\rho) =  {\rm const.} \in {\mathbb R}^+$ and $c_{n\ell}(\rho)=0$ for $(\ell,n) \ne (0,0)$.
	
	Knowing the distribution function allows one to calculate the energy-momentum tensor $T^{\mu \nu}  :=  \left\langle p^{\mu} p^{\nu} \right\rangle_{f}$. For the ansatz~\eqref{eq:boltz_Gub}, the non-zero components are found to only depend on the moments $c_{00},c_{01}$ as follows
	\begin{subequations}
		\begin{align}
		T^{\rho \rho}(\rho) &= \frac{3 {T}(\rho)^4\,c_{00}(\rho)}{\pi^2} , \\
		T^{\theta \theta}(\rho) &= \frac{{T}(\rho)^4}{\pi^2 \cosh^2 \rho} \left[ 1 - \frac{1}{5}c_{01}(\rho) \right], \\
		T^{\phi \phi}(\rho,\theta) &= \frac{{T}(\rho)^4}{\pi^2 \cosh^2 \rho \sin^2 \theta} \left[ 1 - \frac{1}{5} c_{01}(\rho) \right], \\
		T^{\varsigma\varsigma}(\rho) &= \frac{ {T}(\rho)^4}{\pi^2} \left[ 1 + \frac{2}{5} c_{01}(\rho) \right],
		\end{align}
	\end{subequations}
	where the Landau matching condition forces $c_{00}(\rho)=1$. 
	It is then easy to identify the hydrodynamic quantities from each component of the energy-momentum tensor as
	\be
	\label{eq:EMTcomp}
	{\cal E}(\rho):=\frac{3 T(\rho)^{4}}{\pi^2},   \quad  {\cal P}(\rho):= \frac{ T(\rho)^{4}}{\pi^2} = \frac{{\cal E(\rho)}}{3}, \quad \pi^{\varsigma\varsigma}(\rho) = \frac{2T(\rho)^{4} c_{01}(\rho)}{5 \pi^2} ,
	\ee
	where ${\cal E}(\rho)$, ${\cal P}(\rho)$, and ${\pi}^{\varsigma\varsigma}(\rho)$ are the energy density, pressure, shear-stress viscous tensor, respectively. Notice that the energy-momentum tensor is scale invariant, thus $T^{\mu}_{\ \mu}(\rho)=0$. Using this argument, Eqs.~\eqref{eq:EMTcomp} together with the conservation law $\nabla_\mu T^{\mu\nu}=0$ lead to the following differential equation for the temperature
	\be
	\frac{d T}{d \rho} = - \frac{T}{3} \left( 2 - \frac{c_{01}}{10} \right) \tanh \rho . \label{eq:ODE_T_gub}
	\ee
	The evolution equations for the moments $c_{n \ell}(\rho)$ are derived by following a similar procedure outlined in Refs.~\cite{Behtash:2018moe,Behtash:2019txb}. These are given by
	\be
	\frac{ dc_{n\ell}}{d \rho} &=& -  \frac{{T}}{\theta_0}c_{n\ell}  
	-  \left[ \frac{(n+4)c_{01}}{30} c_{n \ell} +   {\frak A}_{n\ell} c_{n\ell+1} + \bar{\frak B}_{n\ell} c_{n\ell}  + {\frak C}_{n\ell} c_{n\ell-1} \right] \tanh \rho  \nl
	&&  + n  \left[ \frac{c_{01}}{30} c_{n-1 \ell} + {\frak D}_{\ell} c_{n-1\ell+1} + \bar{\frak E}_{\ell} c_{n-1 \ell}  + {\frak F}_{\ell} c_{n-1\ell-1} \right] \tanh \rho , \label{eq:ODE_Gub_n_c} 
	\ee
	where
	\be
	{\frak A}_{n\ell} &=& 
	\frac{(2\ell+1)(2\ell+2)(2\ell-n-1)}{(4\ell+3)(4\ell+5)}  , \\
	\bar{\frak B}_{n\ell} 
	&=& -\frac{2 \ell (2 \ell+1) (2 n+5)}{3 (4 \ell -1) (4 \ell + 3)}, \\
	{\frak C}_{n\ell} &=& -\frac{2\ell (2\ell-1)(2\ell+n+2)}{(4\ell-3)(4\ell-1)}, \\
	\bar{\frak E}_\ell
	&=&  -  \frac{4 \ell( 2\ell + 1) }{3(4\ell-1)(4\ell+3)}\,,\\
	{\frak D}_\ell &=& - \frac{2(\ell+1)(2\ell+1)}{(4\ell+3)(4\ell+5)}, \\
	{\frak E}_\ell &=& \frac{1}{{T}} \frac{d{T}}{d \rho} \frac{1}{\tanh \rho} + \frac{2(4\ell^2+2\ell-1)}{(4\ell-1)(4\ell+3)},   \\
	{\frak F}_\ell &=& -\frac{2\ell(2\ell-1)}{(4 \ell-3) (4\ell-1)} .
	\ee
	Eqs. \eqref{eq:ODE_T_gub} as well as \eqref{eq:ODE_Gub_n_c} make up together what we will refer to as the {\it Gubser dynamical system}. 
	
	Practically speaking, it is not possible to solve the Gubser dynamical system in its full-fledged form, which is basically an infinite-dimensional system of coupled differential equations. But one can solve the system systematically with the truncation of the upper bounds $\ell$ and $n$ as $\ell \le  L \in {\mathbb N}, n \le N \in {\mathbb N}_0$ in the ansatz~\eqref{eq:boltz_Gub}. A surprising aspect of Gubser dynamical system is that even though it does
	admit a formal transseries solution, but there are no exponential transmonomials in contrast to the transseries solution of the Bjorken dynamical system. Therefore, the theory is completely perturbative. However, in both systems the evolution of the energy momentum tensor is fully determined by the moments $c_{0l}:=c_l$~\cite{Behtash:2018moe,Behtash:2019txb}. For this reason, we narrow down our search for the transseries structure and properties of the dynamical sub-system 
	\be
	\frac{ dc_{\ell}}{d \rho}
	&=& -\left[ \frac{{T}}{\theta_0}c_\ell  +  \left( \frac{2c_1}{15}c_{\ell} +  {\frak A}_{\ell} c_{\ell+1} + \bar{\frak B}_{\ell} c_{\ell}  + {\frak C}_{\ell} c_{\ell-1} \right) \tanh \rho \right] 
	, \label{eq:ODE_Gub_n0_c} \\
	\frac{ d {T}}{d \rho} &=& -\frac{{T}}{3} \left( 2 - \frac{c_1}{10} \right)\tanh \rho 
	, \label{eq:ODE_Gub_n0_T} 
	\ee
	where the coefficients are given by
	\be
	{\frak A}_{\ell} &=& \frac{(2\ell-1)(2\ell+1)(2\ell+2)}{(4\ell+3)(4\ell+5)}, \label{eq:coeff_A}  \\
	\bar{{\frak B}}_{\ell} &=&   -\frac{10 \ell (2 \ell+1)}{3 (4 \ell-1) (4 \ell+3)}, \label{eq:coeff_bB} \\
	{\frak C}_{\ell} &=& -\frac{4 \ell (2\ell-1)(\ell+1)}{(4\ell-3)(4\ell-1)}. \label{eq:coeff_C}
	\ee 
	\subsection{Flow structure} \label{sec:dyn_gubser}
	
	In this subsection, we investigate the local and global flow structure of the dynamical system (\ref{eq:ODE_Gub_n0_c}) and (\ref{eq:ODE_Gub_n0_T}).
	
	\subsubsection{Fixed point and stability} \label{sec:fixed_Gubser}
	Similarly to the IS theory studied previously, we take $\tau$ to be a variable of flow time $\rho$ in the nonautonomous system~\eqref{eq:ODE_Gub_n0_c} 
	and \eqref{eq:ODE_Gub_n0_T}, and introduce an ODE for $\tau$, which results in the autonomous dynamical system
	\be
	\frac{ dc_{\ell}}{d \rho}
	&=& -  \frac{{T}}{\theta_0}c_\ell  -  \left( \frac{2c_1}{15}c_{\ell} +  {\frak A}_{\ell} c_{\ell+1} + \bar{\frak B}_{\ell} c_{\ell}  + {\frak C}_{\ell} c_{\ell-1} \right) \tau  \, =: \, F_{\ell}({\bf C}) , \label{eq:ODE_Gub_n0_c_auto} \\
	\frac{ d {T}}{d \rho} &=& -\frac{{T}}{3} \left( 2 - \frac{c_1}{10} \right)\tau \, =: \, F_{T} ({\bf C}) , \label{eq:ODE_Gub_n0_T_auto} \\
	\frac{d \tau}{d\rho} &=& 1-\tau^2  \, =:\, F_{\tau}(\tau). \label{eq:ODE_Gub_n0_tau_auto} 
	\ee
	Here, we have defined the vector $\bf C$ whose components stand for the variables of the system, i.e., $C_{i}:=(c_1,\cdots,c_{L},T,\tau)$ where $1 \le i \le L+2$. This vector essentially unveils the structure of configuration space at $\tau$. The general solution of Eq.~(\ref{eq:ODE_Gub_n0_tau_auto})
	is given by $\tau(\rho) = \tanh (\rho - \rho_0)$, where $\rho_0$ is the integration constant. Evidently, the original nonautonomous dynamical system is  reproduced by setting $\rho_0=0$. 
	
\begin{figure}[htbp]
		\begin{center}
			\begin{tabular}{cc}
				\begin{minipage}{0.5\hsize}
					\begin{center}
						\includegraphics[clip, width=70mm]{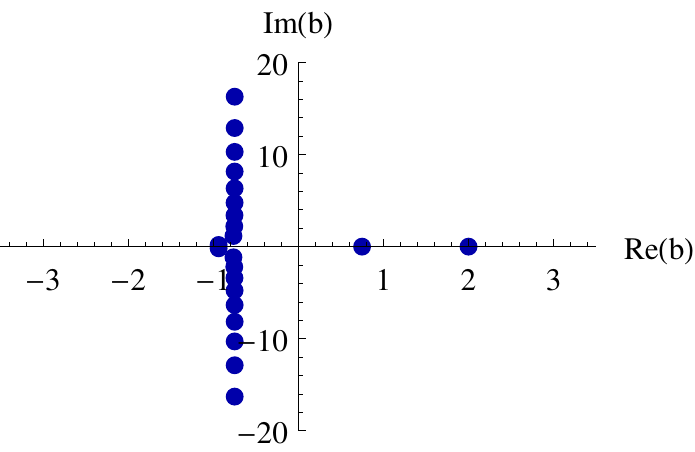}
						\hspace{1.6cm} (a) UV fixed points for $L=20$
					\end{center}
				\end{minipage}
				\begin{minipage}{0.5\hsize}
					\begin{center}
						\includegraphics[clip, width=70mm]{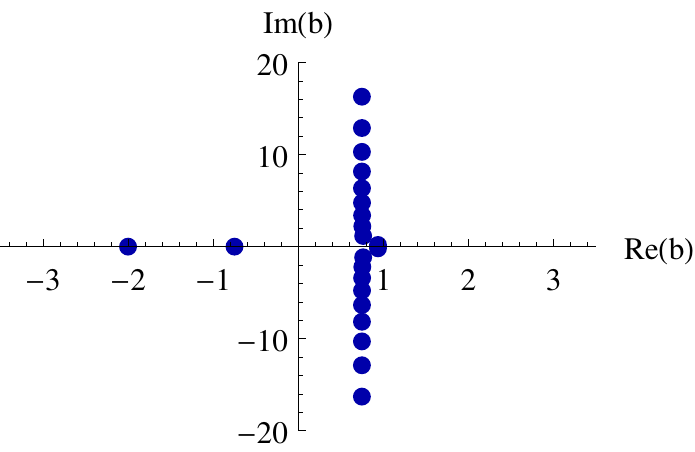}
						\hspace{1.6cm} (b) IR fixed points for $L=20$
					\end{center}
				\end{minipage} \\
				\begin{minipage}{0.5\hsize}
					\begin{center}
						\includegraphics[clip, width=70mm]{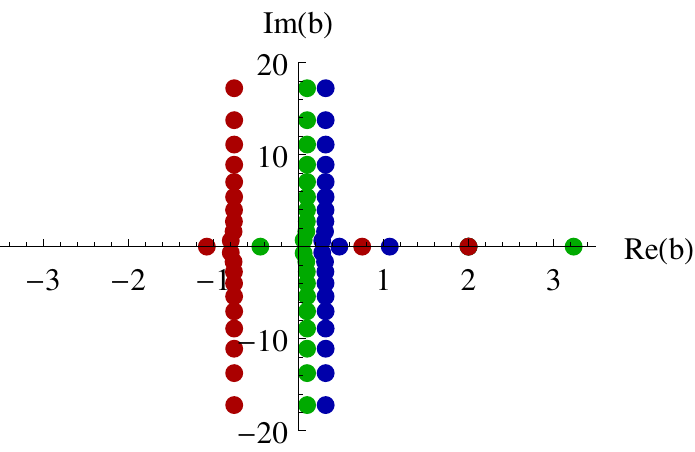}
						\hspace{1.6cm} (c) UV fixed points for $L=21$
					\end{center}
				\end{minipage}
				\begin{minipage}{0.5\hsize}
					\begin{center}
						\includegraphics[clip, width=70mm]{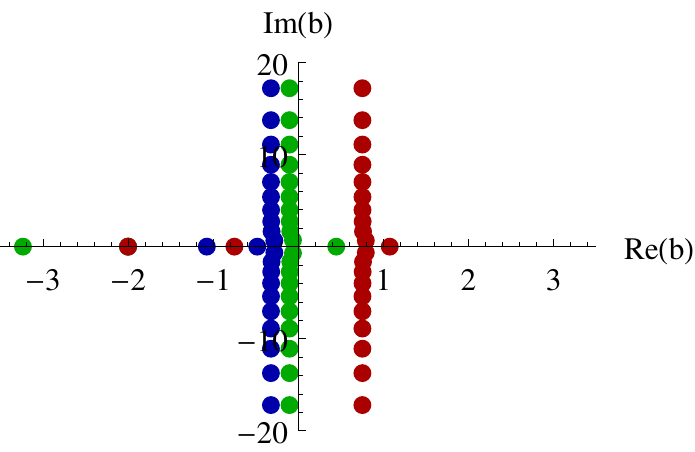}
						\hspace{1.6cm} (d) IR fixed points for $L=21$
					\end{center}
				\end{minipage} 
			\end{tabular} 
			\caption{Plots of the Lyapunov exponents associated with  each fixed point in the case of $L=20$ and $L=21$.
				In (c) and (d), the colored points represent the exponents for $\Sigma_{{\rm U,I(A)}}$(Blue), $\Sigma_{{\rm U,I(B)}}$(Red), and $\Sigma_{{\rm U,I(C)}}$(Green).}
			\label{fig:stability}
		\end{center}
	\end{figure}

	Let $\bar{C}_{i}:=(\bar{c}_{1},\cdots,\bar{c}_{L},\bar{T},\bar{\tau} )$ be a fixed point of the above dynamical system, satisfying $d\bar{ {\bf C}}/d \rho = 0$, i.e.,
	\be
	&&  \frac{\bar{T}}{\theta_0} \bar{c}_\ell  + \left( \frac{2\bar{c}_1}{15} \bar{c}_{\ell} + {\frak A}_{\ell} \bar{c}_{\ell+1} + \bar{\frak B}_{\ell} \bar{c}_{\ell}  + {\frak C}_{\ell} \bar{c}_{\ell-1} \right) \bar{\tau}  = 0 , \label{eq:ODE_Gub_n0_c_fix} \\
	&&   \frac{\bar{T}}{3} \left( 2 - \frac{\bar{c}_1}{10} \right) \bar{\tau} = 0, \label{eq:ODE_Gub_n0_T_fix} \\
	&& 1 - \bar{\tau}^2 = 0. \label{eq:ODE_Gub_n0_tau_fix} 
	\ee
	The sign of the solution of Eq.~(\ref{eq:ODE_Gub_n0_tau_fix}) given by $\lim_{\rho \rightarrow \mp \infty}\tanh \rho=\pm 1$, or equivalently, $\bar{\tau} = \mp 1$ is associated with the UV($-1$) and IR($+1$) limits, respectively. The general solutions are complex, but we can numerically obtain two real solutions a couple of UV and IR fixed points if $L \in 2 {\mathbb N}$, and six real solutions (three UV and three IR fixed points) for $L \in 2{\mathbb N}-1$. For example, the fixed points for $L=20, 21$ are given by
	\be
	&& \bar{\bf C} =
	(-2.47792, 3.50714, -3.84586,  \cdots ,0, \mp 1) \quad =: {\Sigma}_{\rm U,I}, \quad \mbox{for } \  L=20, \\ \nl
	&& \bar{\bf C} = 
	\begin{cases}
		(5.53822,6.37524,15.0072, \cdots, 0 , \mp 1) & =: \Sigma_{\rm U,I(A)} \\
		(-2.51978,3.25608,-4.25828, \cdots, 0,\mp 1) & =: \Sigma_{\rm U,I(B)} \\
		(20, 9.34538 , 64.0704, \cdots, \pm 2.14623, \mp 1) & =: \Sigma_{\rm U,I(C)} 
	\end{cases}, \quad \mbox{for } \ L=21. \label{eq:fixed_C} \nl
	\ee
	The stability of the dynamical system at every fixed point is demonstrated by calculating the Lyapunov exponents or simply the eigenvalues of the Jacobian matrix
	\be
	J^{a}_{ij} = \left. \frac{\pd F_{i}}{\pd C_{j}} \right|_{{\bf C} \rightarrow \Sigma_a},
	\ee
	where $a$ stands for the fixed point $a$. We then find the instability index for each fixed point to be
	\be
	&&   {\rm Ind}[J^{\rm U}] = 2,  \qquad {\rm Ind}[J^{\rm I}] = L, \qquad \mbox{for } \ L \in 2 {\mathbb N}, \\ \nl
	&&   {\rm Ind}[J^{\rm U(A)}] = L+2,  \qquad {\rm Ind}[J^{\rm I(A)}] = 0, \nl
	&&   {\rm Ind}[J^{\rm U(B)}] = 2,  \qquad \qquad {\rm Ind}[J^{\rm I(B)}] = L , \qquad \mbox{for } \ L \in 2{\mathbb N} - 1 , \\
	&&   {\rm Ind}[J^{\rm U(C)}] = L+1,  \qquad {\rm Ind}[J^{\rm I(C)}] = 1. \nn
	\ee
	Fig. \ref{fig:stability} shows the distribution of ${\rm Eigen}(J^{a})=(b_1,\dots,b_{L+2})$ on the complex plane.
	One can see that both the UV and IR fixed points in the truncated system for $L \in 2 {\mathbb N}$ are unstable.
	Therefore, even $L$ does not contain an invariant manifold for the flow space and would not reproduce a physical dynamical system as
	$L\rightarrow\infty$. A similar conclusion was drawn for the truncation of Bjorken dynamical system in \cite{Behtash:2019txb}~\footnote{In order to reproduce the
exact solution, we have to take the limit $L \to\infty$ but how to take this
limit is not always unique. For instance, in our analytical and numerical studies~\cite{Behtash:2019txb}, it was observed that the shear viscous tensor calculated from the exact solution presents two
UV fixed points, each of them located where either the transverse or the longitudinal
pressure vanishes. These two UV fixed points exist for both odd and even number of
moments when truncating the nonlinear ODEs. But if this number is even, there are
only complex solutions in the UV and thus, the UV physics of the exact solution of
the Boltzmann equation is not described by the truncated system in this case. One concludes that it is necessary to take the large $L$ limit as $L = 2n + 1$ with $n \to\infty$.}. In contrast,
	the case of odd $L$ stands out in that we have a variety of different fixed points just as in the IS theory, with $\Sigma_{\rm U(A)}$ being a spiral source, $\Sigma_{\rm I(A)}$ being a spiral sink, and rest of fixed points being saddle points. Hence, a closed invariant manifold might be found in the flow space. For this reason, we always assume hereafter that $L$ is odd.
	
	By calculating the expansion in \eqref{eq:boltz_Gub} at the fixed point (\ref{eq:fixed_C}), one can immediately deduce that
	the Gubser flow does not reach a thermal equilibrium point in the IR limit,
	\be
	\lim_{\rho \rightarrow + \infty} \frac{f(\rho,p_\Omega^2,|p_\varsigma|)}{ f_{\rm eq}(p_{\rho}/T(\rho))}    \ne 1 .
	\ee
	However, the distribution function correctly converges to a steady state because our collisional kernel ${\cal C}[f]$ is proportional to ${T}(\rho)$, and thus, 
	\be
	\lim_{\rho \rightarrow +\infty} {T}(\rho) = 0 \quad &\Rightarrow& \quad \lim_{\rho \rightarrow +\infty}  \mathcal{C}[f] =0.
	\ee
	This result will remain the same for any $L$ and $N$. 
	
	\subsubsection{Global structure}
	We consider the global flow structure of the system in Eqs.~(\ref{eq:ODE_Gub_n0_c}) and (\ref{eq:ODE_Gub_n0_T}). The analysis is similar  to the what was done for the IS theory in Sec.~\ref{sec:global_IS}. Since the results will also be similar to those of IS theory, we are going to avoid divulging the technicalities and results and rather focus on the conclusions and main differences distinguishing the dynamical current system from IS theory. Moving forward, we will again assume $L \in 2 {\mathbb N}-1$ for obtaining a source (to mimic the high energetic streaming of particles) and a sink (non-thermal equilibrium) in the UV and IR limits respectively, as explained in Sec.~\ref{sec:fixed_Gubser}.
	The following is a quick sketch of the setup and what needs to be understood from the resulting flow analysis.
	\begin{enumerate}
		\item
		The vector bundle $\Theta=({\cal M},\pi,B,F)$ is given by
		\be
		\mbox{Total space} &:& {\cal M} = B \times F, \\
		\mbox{Projection} &:& {\pi} : {\cal M} \rightarrow B, \\
		\mbox{Base space} &:& B = I := (-1,+1) \ni \tau, \\
		\mbox{Fiber space} &:& F = {\mathbb R}^{L+1} \ni {\bf c} = (c_{1},\cdots,c_{L},T).
		\ee
		The nonautonomous system $(\theta,\phi)$ is given by the solution of 
		\be
		\frac{ dc_{\ell}}{d \rho}
		&=& -  \frac{{T}}{\theta_0}c_\ell  -  \left( \frac{2c_1}{15}c_{\ell} +  {\frak A}_{\ell} c_{\ell+1} + \bar{\frak B}_{\ell} c_{\ell}  + {\frak C}_{\ell} c_{\ell-1} \right) \tau  \, =: \, F_{\ell}({\bf c};\tau) , \label{eq:ODE_Gub_n0_c_auto2} \\
		\frac{ d {T}}{d \rho} &=& -\frac{{T}}{3} \left( 2 - \frac{c_1}{10} \right)\tau \, =: \, F_{T} ({\bf c};\tau) , \label{eq:ODE_Gub_n0_T_auto2} 
		\ee
		in which we set $\tau(\rho) = \tanh \rho$ for $\theta_\rho$.
		\item
		One can obtain the fixed point $\bar{\bf c}$ on the $\tau$-fiber by solving the fixed point equation defined as $F_{i} (\bar{\bf c};\tau)= 0$.
		One can have two real constant solutions: (1) $\Sigma_{\rm A,B}$ corresponding to $\Sigma_{\rm U,I(A)}$ and $\Sigma_{\rm U,I(B)}$ in Eq.~(\ref{eq:fixed_C}) at $\tau=\mp 1$; and (2) a $\tau$-independent real solution corresponding to $\Sigma_{\rm U,I(C)}$.
		The $\tau$-dependence only enters the $T$-component of $\bar{\bf c}$ linearly, that is $\bar{T}(\tau)=- \tau \bar{A}$ with a positive constant $\bar{A} \approx 2.1$.
		The stability of the three fixed points on ${\cal M}_{\tau}$ changes at $\tau = 0$ in accordance with the 
		symmetry rule
		\be
		{\rm Ind}[J^{{\rm U}(\bullet)}] = L + 1 - {\rm Ind}[J^{{\rm I}(\bullet)}].
		\ee
		\item
		For $L = 2{\mathbb N } -1$, one can define the invariant manifold of the flow space ${\cal S}$ connecting to the fixed points given by Eq.~(\ref{eq:fixed_C}).
		Using ${\rm Diff}({\cal S})$, an identical version of the decomposition done for the IS theory might be achieved here as
		\be
		&& {\cal S} = {\cal S}_{\rm AA} \cup {\cal S}_{\rm AB} \cup {\cal S}_{\rm AC} \cup {\cal S}_{\rm BA} \cup {\cal S}_{\rm BB} \cup {\cal S}_{\rm CA}, \\
		&& {\cal S}_{\rm BC}  = {\cal S}_{\rm BC} = {\cal S}_{\rm CC} = \emptyset. 
		\ee
		Furthermore, from the perspective of the Morse decomposition,
		the global structure of the flow space is unfolded in a seemingly similar way as in discussion of IS theory, giving rise to the following 
		connection diagram
		\be
		\begin{tikzcd}
			{\cal R}_{\rm B}   \arrow[dd]  \arrow[ddr]  &   {\cal R}_{\rm A} \arrow[dd]  \arrow[ldd]  \arrow[l,dashed] \arrow[r,dashed] \arrow[rdd] & {\cal R}_{\rm C}   \arrow[ldd]  \\
			& & \\
			{\cal A}_{\rm B} \arrow[r,dashed] & {\cal A}_{\rm A}  & {\cal A}_{\rm C}  \arrow[l,dashed]
		\end{tikzcd} \label{eq:flow_full2}
		\ee
	\end{enumerate}
	
	\subsection{Transseries construction} \label{sec:trans_gugser}
	In this subsection, we construct the transseries solution expanded around the IR fixed point $\Sigma_{\rm I(A)}$.
	The transseries analysis for the Gubser flow shows that not only it is not possible to interpret hydrodynamic or non-hydrodynamic modes, but also provides yet another evidence for the non-existence of the attractor solution.
	Later on, we also construct the transseries solution around the UV fixed point $\Sigma_{\rm U(A)}$, which can be identically done for all the other fixed points.
	
	For $L \in 2{\mathbb N} -1$ and $N=0$, the ODEs are given by
	\be
	&& \frac{ dc_{\ell}}{d z} = -\frac{1}{2z} \left[ \frac{{T}}{\theta_0}c_\ell +  ( {\frak A}_{\ell} c_{\ell+1} + {\frak B}_{\ell} c_{\ell}  + {\frak C}_{\ell} c_{\ell-1} ) \frac{1-z^{-1}}{1+z^{-1}}  \right] \nl
	&& \quad \, \, \, \, \,  = -\frac{1}{2z} \left[   \frac{{T}}{\theta_0} c_\ell + \left(  \frac{2c_1}{15} c_\ell + {\frak A}_{\ell} c_{\ell+1} + \bar{{\frak B}}_{\ell} c_{\ell}  + {\frak C}_{\ell} c_{\ell-1} \right) \frac{1-z^{-1}}{1+z^{-1}} \right], \\
	&& \frac{ d {T}}{d z} = -\frac{{T}}{6z} \left( 2 - \frac{c_1}{10} \right) \frac{1-z^{-1}}{1+z^{-1}}.
	\ee
	Note that $c_0 = 1$ due to the Landau matching condition. The fixed points at the IR limit can be obtained by solving $dc_\ell/dz = d {T}/dz = 0$ at $z \rightarrow \infty$.
	Let us denote the fixed points by $\bar{\bf c}=(\bar{c}_\ell,\bar{T})$.
	Notice that $\bar{c}_{\ell}$ has an $L$-dependent nonzero value for any $\ell$. In our analysis, we choose $\Sigma_{\rm I(A)}$ to represent the fixed point for which $\bar{T}=0$. Shifting $c_\ell \rightarrow c_\ell + \bar{c}_{\ell}$, we get
	\begin{subequations}
		\label{eq:ODE_Gub}
		\begin{align}
		\frac{dc_\ell}{dz} &= -\frac{1}{2z} \left[  \frac{{T}}{\theta_0} c_\ell +  \frac{\bar{c}_\ell}{\theta_0} {T}  + \left\{      \frac{2c_1}{15} c_\ell + \frac{2\bar{c}_\ell}{15} c_1   \right. \right. \nonumber  \nl
		& \left. \left. + {\frak A}_{\ell} c_{\ell+1} + \left( \bar{{\frak B}}_{\ell} + \frac{2 \bar{c}_1}{15} \right) c_{\ell}  + {\frak C}_{\ell} c_{\ell-1} 
		\right\}  \frac{1-z^{-1}}{1+z^{-1}} \right] , \label{eq:ODE_Gub_c1} \\
		\frac{ d {T}}{d z} &= -\frac{{T}}{6z} \left( A_T - \frac{c_1}{10} \right) \frac{1-z^{-1}}{1+z^{-1}}, \label{eq:ODE_Gub_T}
		\end{align}
	\end{subequations}
	where $A_T = 2 - \bar{c}_1/10$ (the shift $c_\ell \rightarrow c_\ell + \bar{c}_{\ell}$ implies $\bar{c}_0=1$ and $c_0=0$.).
	Next, we define the $(L+1)$-component vector ${\bf c} = (c_1,c_2,\cdots,c_L,{T})^\top$. This will cast~\eqref{eq:ODE_Gub} in the form 
	\be
	(z+1)\frac{d{\bf c}}{dz} = - \left( {\cal A}^{(+)} + z^{-1} {\cal A}^{(-)}\right)  {\bf c} , \label{eq:diff_gub_z_c}
	\ee
	where ${\cal A}^{(\pm)}$ is an $(L+1)$-by-$(L+1)$ matrix, decomposed into a constant part and a ${\bf c}$-dependent part expressed by ${\cal A}^{(\pm)} = {\cal B}^{(\pm)}  + {\cal D}^{(\pm)}({\bf c})$.
	The matrices ${\cal B}^{(\pm)}$ and ${\cal D}^{(\pm)}$ are respectively
	\be 
	&&   {\cal B}^{(\pm)} = \pm \frac{1}{2}
	\begin{pmatrix} 
		\bar{{\frak B}}_1^{\prime} + \frac{2 \bar{c}_1}{15} & {\frak A}_1 & & & & & \pm \frac{\bar{c}_1}{\theta_0} \\
		{\frak C}_2 + \frac{2 \bar{c}_2}{15}     & \bar{{\frak B}}_2^\prime & {\frak A}_2 & & & & \pm \frac{\bar{c}_2}{\theta_0} \\
		\frac{2 \bar{c}_3}{15}  & {\frak C}_3 & \bar{{\frak B}}_3^\prime & {\frak A}_3 & & & \pm \frac{\bar{c}_3}{\theta_0} \\
		\vdots &   & \ddots & \ddots & \ddots & & \vdots \\
		\frac{2 \bar{c}_{L-1}}{15} & & & {\frak C}_{L-1} & \bar{{\frak B}}_{L-1}^\prime & {\frak A}_{L-1} & \pm \frac{\bar{c}_{L-1}}{\theta_0} \\
		\frac{2 \bar{c}_L}{15} & & & & {\frak C}_L & \bar{{\frak B}}_L^\prime  & \pm \frac{\bar{c}_L}{\theta_0} \\
		& & & & & & \frac{A_T}{3}
	\end{pmatrix}, \quad \bar{{\frak B}}_\ell^\prime = \bar{{\frak B}}_\ell + \frac{2 \bar{c}_1}{15}, \\
	&& {\cal D}^{(\pm)}({\bf c}) = \pm \frac{1}{2}
	\begin{pmatrix} 
		\pm\frac{{T}}{\theta_0} + \frac{2c_1}{15} & & &  & \\
		& \pm\frac{{T}}{\theta_0} + \frac{2c_1}{15} & &  & \\
		& & \ddots & &  \\
		& & & \pm\frac{{T}}{\theta_0} + \frac{2c_1}{15} &  \\
		& & & & -\frac{c_1}{30} 
	\end{pmatrix}, 
	\ee
	To simplify the matters, we are going to seek a diagonalized form of ${\cal B}^{(+)}$ by acting on $\bf c$ 
	by an invertible matrix $U$, and following the diagonalization procedure as
	\begin{subequations}
		\begin{align}
		& \tilde{\bf c} = U {\bf c}, \label{eq:Utrans}\\
		& \tilde{\cal A}^{(\pm)} = \tilde{\cal B}^{(\pm)} + \tilde{\cal D}^{(\pm)}({\bf c}) , \\
		& \tilde{\cal B}^{(+)} = U {\cal B}^{(+)} U^{-1} = {\rm diag}(b_1,\cdots,b_L,b_{L+1}), \\
		& \tilde{\cal B}^{(-)} = U {\cal B}^{(-)} U^{-1}, \\
		& \tilde{\cal D}^{(\pm)}({\bf c}) = U{\cal D}^{(\pm)}({\bf c})U^{-1}.
		\end{align}
	\end{subequations}
	In terms of the variables with a tilde, the dynamical system
	takes the form
	\be
	(z+1)\frac{d \tilde{\bf c}}{dz} &=& - \left( \tilde{\cal A}^{(+)} + z^{-1} \tilde{\cal A}^{(-)} \right) \tilde{\bf c} \nl
	&=&  - \left(\tilde{\cal B}^{(+)} + z^{-1} \tilde{\cal B}^{(-)} + \tilde{\cal D}^{(+)}({\bf c}) + z^{-1} \tilde{\cal D}^{(-)}({\bf c}) \right) \tilde{\bf c}. \label{eq:diffeq_gubser}
	\ee
	The formal transseries ansatz to solve these equations is given by
	\be
	&& \tilde{{c}_\ell}(z) = \sum_{k=0}^\infty \sum_{|{\bf m}|\ge 0}^\infty \tilde{u}_{\ell,k}^{({\bf m})} \Phi^{\bf m}_{k}, \qquad \mbox{as \ } z \rightarrow +\infty,  \label{eq:ansatz_Gub} \\
	&& \Phi^{\bf m}_{k}
	:= \bm{\sigma}^{\bf m} z^{{\bf m} \cdot \bm{\beta}-k}, \qquad \bm{\sigma}^{\bf m}:= \prod_{\ell=1}^{L+1}  \sigma_\ell^{m_\ell}
	\ee
	where ${\bf m} \in {\mathbb N}_0^{L+1}$, $\bm{\beta} \in {\mathbb C}^{L+1}$ are the vectorial indices and eigenvalues characterizing the first-order corrections in the transseries, and $\bm{\sigma} \in {\mathbb C}^{L+1}$ are the integration constants.
	Notice that $\tilde{u}_{\ell,0}^{({\bf 0})} = 0$ for any $\ell$.
	In the original variables $(c_\ell,{T})$ of the dynamical system, the form of transseries \eqref{eq:ansatz_Gub} is preserved, with the 
	coefficients being given by the action of $U^{-1}$ on $\tilde{\bf u}^{(\mathbf{m})}_{k}$, namely
	\begin{subequations}
		\begin{align}
		\label{eq:cansatz}
		& c_\ell(z) = \sum_{k=0}^\infty \sum_{|{\bf m}|\ge 0}^\infty u_{\ell,k}^{({\bf m})} \Phi^{\bf m}_{k}, \\
		& u_{\ell,k}^{({\bf m})} = \sum_{\ell^\prime=1}^{L+1} U^{-1}_{\ell \ell^\prime} \tilde{u}_{\ell^\prime,k}^{({\bf m})}.
		\end{align}
	\end{subequations}
	We note that ${\bf u}_{k}^{({\bf m})}$ in Eq.~(\ref{eq:Dnk}) can be calculated similarly using ${\bf u}_{k}^{({\bf m})}=U^{-1} \tilde{\bf u}^{({\bf m})}_k$. Moreover, $\tilde{\cal D}^{(\pm)}({\bf c})$ might be expanded in the same basis as
	\be
	&&   {\cal D}^{(\pm)}_{\ell\ell^\prime}(z)  = 
	\sum_{|{\bf m}|\ge 0}^\infty \sum_{k=0}^\infty 
	{\cal D}_{\ell\ell^{\prime},k}^{(\pm)({\bf m})}  \Phi^{\bf m}_{k}, \\
	&& {\cal D}_{\ell \ell^{\prime},k}^{(\pm)({\bf m})}    = \delta_{\ell, \ell^\prime} \times
	\begin{cases}
		\pm \frac{u_{1,k}^{({\bf m})}}{15} + \frac{u_{L+1,k}^{({\bf m})}}{2\theta_0} 
		& \mbox{for } \ell=1,\cdots,L \\
		\mp \frac{u_{1,k}^{({\bf m})}}{60}  & \mbox{for } \ell=L+1 
	\end{cases}, \label{eq:Dnk}   \\
	&& \tilde{\cal D}^{(\pm)}_{\ell\ell^\prime}(z) = \sum_{\ell_1,\ell_2=1}^{L+1}
	U_{\ell \ell_1} {\cal D}^{(\pm)}_{\ell_1 \ell_2}(z) U^{-1}_{\ell_2 \ell^\prime} \, := \, \sum_{|{\bf m}|\ge 0}^\infty \sum_{k=0}^\infty  \tilde{D}^{(\pm)(\bf n)}_{\ell\ell^\prime,k} \Phi^{\bf m}_{k}. 
	\ee
	
	Substituting the ansatz~\eqref{eq:ansatz_Gub} into Eq.~(\ref{eq:diffeq_gubser}) gives the evolution equation for the coefficients as
	\be
	&& \left( {\bf m}\cdot \bm{\beta} + b_\ell  - k \right) \tilde{u}^{({\bf m})}_{\ell,k} + \left(  {\bf m}\cdot \bm{\beta}  - k  +1\right) \tilde{u}^{({\bf m})}_{\ell,k-1} + \sum_{\ell^\prime=1}^{L+1} \tilde{\cal B}^{(-)}_{\ell\ell^\prime} \tilde{u}^{({\bf m})}_{\ell^\prime,k-1} \nl
	&& +  \sum_{\ell^\prime=1}^{L+1} \sum_{|{\bf m}^\prime| \ge 0}^{{\bf m}} \left( \sum_{k^\prime=0}^{k} \tilde{\cal D}_{\ell\ell^\prime,k-k^\prime}^{(+)({\bf m}-{\bf m}^{\prime})} \tilde{u}^{({\bf m}^{\prime})}_{\ell^\prime,k^\prime} + \sum_{k^\prime=0}^{k-1} \tilde{\cal D}_{\ell\ell^\prime,k-k^\prime-1}^{(-)({\bf m}-{\bf m}^{\prime})} \tilde{u}^{({\bf m}^{\prime})}_{\ell^\prime,k^\prime} \right) = 0. \label{eq:evo_gubser}
	\ee
	We can further fix $\tilde{u}^{({\bf m})}_{\ell,0}=1$ for $m_{\ell^\prime}=\delta_{\ell,\ell^{\prime}}$ to normalize the integration constants. Eq.~\eqref{eq:evo_gubser} is then solved for all the unknown coefficients $\tilde{u}^{({\bf m})}_{\ell,k}$ and $\beta_\ell$  unambiguously. It is straightforward to show that
	\be
	&& \tilde{u}^{({\bf 0})}_{\ell,k}  = 0 \quad \mbox{for any $\ell$ and $k$}.
	\ee
	and $\beta_\ell = - b_\ell$. After determining $\tilde{u}^{({\bf m})}_{\ell,k}$ and $\beta_{\ell}$, the formal transseries solution is written as
	\be
	\label{eq:transsol_c}
	{\bf c}(z) = \sum_{|{\bf m}|>0}^{+\infty} \sum_{k=0}^{+\infty} {\bf u}^{({\bf m})}_{k} \Phi^{\bf m}_{k} + \bar{\bf c}, 
	\ee
	where ${\bf u}^{{({\bf m})}}_k = U^{-1} \tilde{\bf u}^{{({\bf m})}}_k$ and $\bar{\bf c}$ is the constant part appearing due to the shift implemented earlier. Comparing Eq.~\eqref{eq:diffeq_gubser} against the prepared form of 
	a generic multi-dimensional rank-1, level-1 vector differential equation around a fixed point at infinity given in
	\cite{10.1155/S1073792895000286}, we are missing the term proportional to $\tilde{\bf c}$ due to the factor $(z+1)$
	on the l.h.s. of that equation. This means that the system does not accept any further exponential corrections so
	the formal solution in Eq.~\eqref{eq:transsol_c} is merely a perturbative series, and in fact a convergent one \footnote{The existence of a transseries solution of
		the dynamical system written fully in the variable $z$ is enough to have a definite conclusion just on the status of convergence, no matter what the radius of convergence would be. Had we started off with $\rho$ instead, this
		conclusion would have not been straightforward to draw.}.
	
	Following the construction of transseries solutions for the IS theory, the formal transseries around the UV fixed point $\Sigma_{\rm U(A)}$ is generated by applying $z \rightarrow \tilde{z}:=1/z$, ${\cal A}^{(+)} \rightarrow {\cal A}^{(-)}$ in the system of (\ref{eq:diff_gub_z_c}), making the UV limit be defined as $\tilde{z} \rightarrow +\infty$. Repeating the analysis leading to 
	Eq.~\eqref{eq:transsol_c}, one can calculate the early-time formal transseries. This might well be achieved by simply following an isomorohism between the transseries expanded around the UV and IR fixed points, of the form
	${\mathbb C}[ z^{-1}, \sigma_1 z^{\beta_1}, \cdots ,\sigma_{L+1} z^{\beta_{L+1}}] \rightarrow {\mathbb C}[ \tilde{z}^{-1}, \sigma_1 \tilde{z}^{\beta_1}, \cdots ,\sigma_{L+1} \tilde{z}^{\beta_{L+1}}]$, and replacing $\theta_0 \rightarrow - \theta_0$.
	
	\subsection{Stability analysis when $N>0,L>0$}
	
	The Gubser dynamical system has a chain structure, i.e., the solution for the moment $c_{nl}$ depends on the higher-order moments $c_{n+1,l+1}$. The moments in the $N=0$ sector only couple with the moments in the same sector, and one can express the energy-momentum tensor in terms of a linear combination of them. The higher moments in the $N>0$ sector are known to express the energy tail for the distribution function~\cite{Behtash:2018moe,Behtash:2019txb}. Therefore, the stability analysis of the system including $N>0$ sectors in both UV and IR limits gives a global picture of the nonlinear relaxation processes of the Gubser dynamical system.
	
	Suppose that we have a system $l<L,\, n<N$, the total dimension of the configuration space is $I+2$, where $I=(N+1)(L+1)-1$. Again, the stability analysis can unfold the instability index for each fixed point. Following the same procedure outlined for the IS theory
	and the sector $N=0$, we find that ${\rm Ind}[J^{\rm U}]$ around all the fixed points are 
	\be
	&&   {\rm Ind}[J^{\rm U}] = N+2,  \qquad {\rm Ind}[J^{\rm I}] = I-N, \qquad \mbox{for } \ L \in 2 {\mathbb N}, \\ \nl
	&&   {\rm Ind}[J^{\rm U(A)}] = I-N+2,  \qquad {\rm Ind}[J^{\rm I(A)}] = N, \nl
	&&   {\rm Ind}[J^{\rm U(B)}] = N+2,  \qquad \qquad {\rm Ind}[J^{\rm I(B)}] = I-N, \qquad \mbox{for } \ L \in 2{\mathbb N} - 1 , \\
	&&   {\rm Ind}[J^{\rm U(C)}] = I-N+1,  \qquad {\rm Ind}[J^{\rm I(C)}] = N+1. \nn
	\ee
	
	These values seem to dismiss the existence of pullback(forward) attractors in the UV(IR) limit.
	However, by reducing the dimension of the fiber space and choosing particular initial conditions for the flows, one can find such a pullback(forward) attractor. This can also be achieved by constraining the integration constants of the transseries solutions to
	the dynamical system around both UV and IR fixed points.
	
	\begin{figure}
		\centering
		\includegraphics[width=1.0\textwidth]{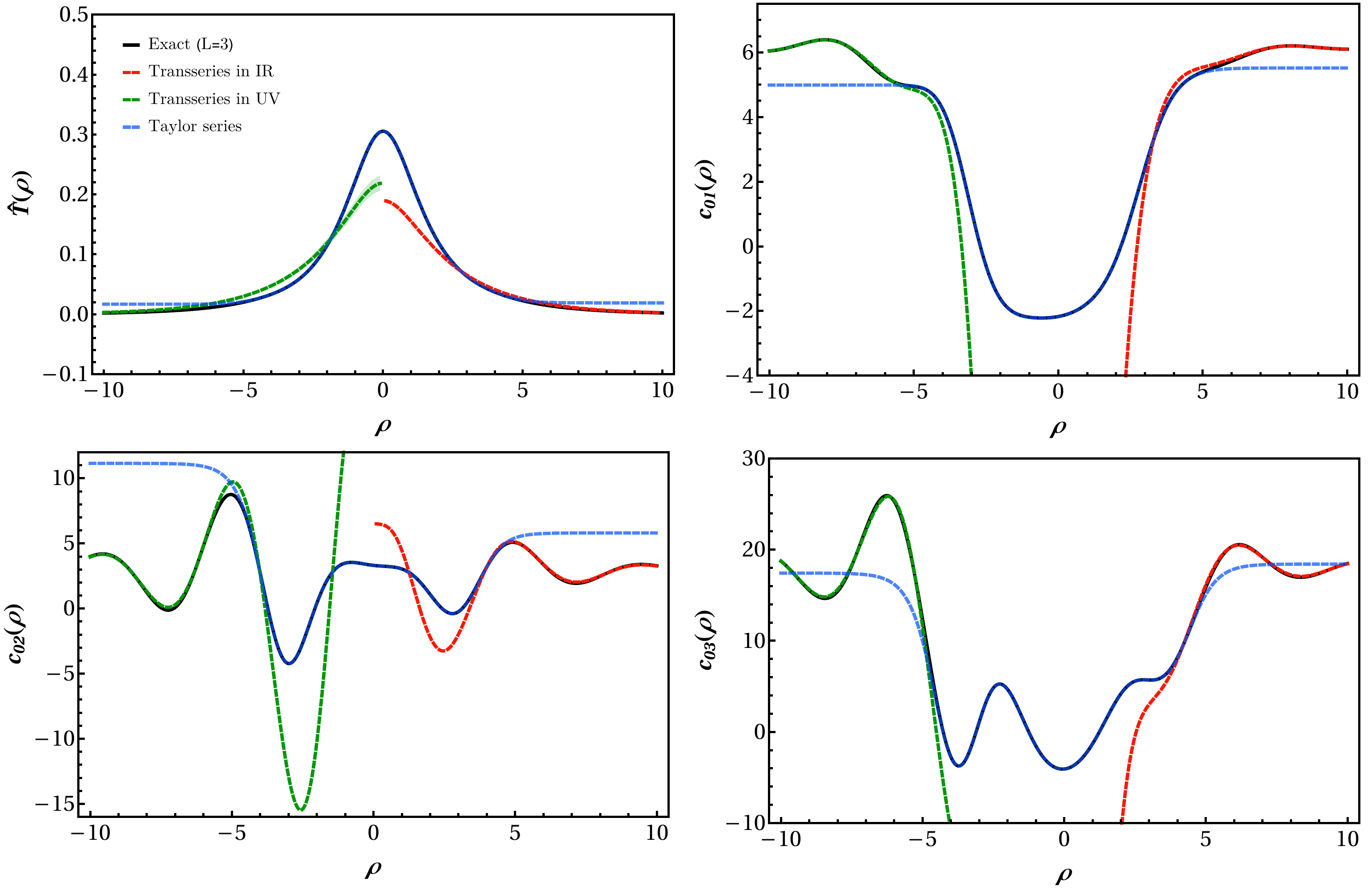}
		\caption{Construction of transseries around the UV fixed point (green dashed line) and IR fixed point (red dashed line) compared to the exact numerical solution (black). The exact solution is computed by setting $\theta_0=25/2\pi,\, c_{01}(\rho_0)=6.1744,\, c_{02}(\rho_0)=1.6959,\, c_{03}(\rho_0)=17.492,\, T(\rho_0)=0.0014$, where the initial time is set to $\rho_0=-10$. The transseries is computed by using first-order non-perturbative corrections and including up to 30th-order perturbative terms around the fixed points. Their integration constants are calculated by simultaneous least-square data-fitting technique using the data of exact solution. Moreover, we have computed the Taylor expansion at $\rho=0$ up to 10000th order.}
		\label{fig:compare}
	\end{figure}
	
	We conclude this section by presenting in Fig.~\ref{fig:compare} a comparison between the numerical solutions of the truncated dynamical system (with $N=0$, $L=3$) (black lines), multiparameter transseries in the UV (green line), IR (red line) and a Taylor series expansion  around $\rho=0$ up to 10000th order (blue line). The discrepancy in the $\rho\to\pm\infty$ regime is expected between exact and transseries
	since $z\rightarrow 1$ as $\rho\rightarrow0$. It is also observed that there is a good agreement between the Taylor series expansion at $\rho=0$ and the exact result which naturally diverges at large $|\rho|$ due to a finite radius of convergence. In all these cases, it is always possible to analytically continue the solutions beyond the original finite domain.
	
	\section{Comparison with the Bjorken flow} \label{sec:com_bjorken}
	Even though the Gubser flow manifests scale invariance and originates from the conformal subgroup of the isometry of the $AdS_5$ spacetime~\cite{Gubser:2010ze,Gubser:2010ui}, its global flow structure and formal transseries are quite different from those of Bjorken flow within the kinetic theory framework~\cite{Behtash:2018moe,Behtash:2019txb}.
	
	In Bjorken flow, the distribution function is invariant under $ISO(2) \times SO(1,1) \times {\mathbb Z}_2$ symmetry
	and one can derive the evolution equations of the moments in the RTA approximation using the moment method with the orthogonal polynomial functions identical to the ones used in the ansatz~\eqref{eq:ansatz_Gub} (see for instance Eq.~(2) in Ref.~\cite{Behtash:2018moe}).
	For $N=0$, the Bjorken dynamical system is given by~\cite{Behtash:2018moe,Behtash:2019txb} 
	\be
	&& \frac{d c_{\ell}}{d \tau} = -  \frac{T}{\theta_0}c_{\ell}  + \frac{1}{\tau} \left( \frac{2 c_{1}}{15}c_{\ell} + {\frak A}_{\ell} c_{\ell+1} + \bar{\frak B}_{\ell} c_{\ell} + {\frak C}_{\ell} c_{\ell-1} \right) , \label{eq:Bjorken_c} \\
	&& \frac{d T}{d \tau} = - \frac{T}{3 \tau} \left( 1 + \frac{c_{1}}{10} \right), \label{eq:Bjorken_T} 
	\ee
	where the coefficients ${\frak A}_{\ell},\bar{\frak B}_{\ell},{\frak C}_{\ell}$ are given by Eqs.~(\ref{eq:coeff_A})-(\ref{eq:coeff_C}) and $\tau \in {\mathbb R}^+$ is the Milne time. The UV and IR limits are described by $\tau \rightarrow 0$ and $\tau \rightarrow +\infty$, respectively.
	
	The global structure of the Bjorken dynamical system can be investigated by following the same procedure described in the previous sections. For example, if one takes $L=21$, the UV/IR fixed points $\bar{\bf c}=( \bar{c}_1, \cdots \bar{c}_{L},\bar{T})$ are
	\be
	\bar{\bf c} = 
	\begin{cases}
		(5.53822,6.37524,15.0072, \cdots, 0 ) & =: \Sigma_{\rm U,I(A)} \\
		(-2.51978,3.25608,-4.25828, \cdots, 0 ) & =: \Sigma_{\rm U,I(B)} \\
		(0, \cdots, 0 ) & =: \Sigma_{\rm I(C)} \\
	\end{cases}. \label{eq:fixed_Bjor} 
	\ee
	In this nonautonomous system, we can obtain two disconnected regions of the flow, and those are characterized by either
	(i) $T(\tau) = 0$ for all $\tau$; or (ii) $T(\tau) \ne 0$.
	If one takes $T(\tau)=0$ for all $\tau$, the ODEs of (\ref{eq:Bjorken_c}) reduce to
	\be
	&& \frac{d c_{\ell}}{d \log \tau} =    \frac{2 c_{1}}{15}c_{\ell} + {\frak A}_{\ell} c_{\ell+1} + \bar{\frak B}_{\ell} c_{\ell} + {\frak C}_{\ell} c_{\ell-1}, \label{eq:Bjorken_c_T0}
	\ee
	and thus, which is essentially an autonomous system. In generic autonomous systems, the flow structure such as the location of fixed points and their stability do not depend on the flow time. In the case of Eq.~(\ref{eq:Bjorken_c_T0}), $\Sigma_{\rm U,I(A)}$ and $\Sigma_{\rm U,I(B)}$ are a source and a sink, respectively, so that there exist three types solutions
	\footnote{
		In an autonomous system, the smallest ingredient of the Morse decomposition must be a fixed point, unlike the nonautonomous systems}
	\be
	\begin{tikzcd}
		{\Sigma}_{\rm U(A)} \arrow[rrdd]  \arrow[dd] &    &  {\Sigma}_{\rm U(B)}  \arrow[dd] \\
		& & \\
		\Sigma_{\rm I(A)} & & \Sigma_{\rm I(B)} 
	\end{tikzcd} \label{eq:flow_bj}
	\ee
	where $\Sigma_{\rm U(A)} \rightarrow \Sigma_{\rm I(A)}$  and $\Sigma_{\rm U(B)} \rightarrow \Sigma_{\rm I(B)}$ are nothing but the trivial solutions exactly at the fixed points, and only $\Sigma_{\rm U(A)} \rightarrow \Sigma_{\rm I(B)}$ gives a nontrivial flow line.
	Unlike a nonoutonomous system, the fixed points $\Sigma_{\rm A,B}=\Sigma_{\rm U,I(A,B)}$ themselves are now invariant submanifolds. The AR pair such as $(\Sigma_{\rm B},\Sigma_{\rm A})$ is then associated with an invariant manifold that is closed and disconnected from $\Sigma_{\rm I(C)}$.
	
	When turning on the temperature, however, the ODEs (\ref{eq:Bjorken_c}) and (\ref{eq:Bjorken_T}) describe a nonautonomous system. In this case, there is only one IR fixed point $\Sigma_{\rm I(C)}$ for the flows, while there exists no UV fixed point because the temperature increases as we approach the UV limit (its time derivative never settles at $0$).
	Since all the components of $\bar{c}_{\ell}$ vanish at $\Sigma_{\rm I(C)}$, the distribution converges to the thermal equilibrium and all the flows reach $\Sigma_{\rm I(C)}$.
	If one takes $w:=\tau T(\tau) \in {\mathbb R}^+_0$ for the flow time, the following dynamical system is derived~\cite{Behtash:2019txb,Behtash:2018moe}
	\be
	\frac{d c_{\ell}}{d w}  = -  \frac{3}{2(1-\frac{c_{1}}{20})} \left[ \frac{c_{\ell}}{\theta_0}  - \frac{1}{w} \left( \frac{2 c_{1}}{15}c_{\ell} + {\frak A}_{\ell} c_{\ell+1} + \bar{\frak B}_{\ell} c_{\ell} + {\frak C}_{\ell} c_{\ell-1} \right) \right] . \label{eq:Bjorken_cw}
	\ee
	This has two UV fixed points $\Sigma_{\rm U(A,B)}$ and one IR fixed point $\Sigma_{\rm I(C)}$, which are exactly the same as those of the original ODEs (\ref{eq:Bjorken_c}) and (\ref{eq:Bjorken_T}). Because $\Sigma_{\rm U(A)}$, $\Sigma_{\rm U(B)}$ and $\Sigma_{\rm I(C)}$ are a source, a saddle, and a sink, respectively, the maximal invariant submanifold ${\cal S}$ might be decomposed into
	\be
	{\cal S} = {\cal S}_{\rm AC} \cup {\cal S}_{\rm BC}.
	\ee
	Indeed, $\Sigma_{\rm U(A)}$  and $\Sigma_{\rm U(B)}$ can be regarded as an end point of a local forward and pullback attractors, respectively,
	and the Morse decomposition is given by
	\be
	\begin{tikzcd}
		{\cal R}_{\rm A} \arrow[rrdd]  \arrow[rr,dashed] &    &  {\cal R}_{\rm B}  \arrow[dd] \\
		& & \\
		& & {\cal A}_{\rm C} 
	\end{tikzcd} \label{eq:flow_bjw}
	\ee
	The AR pair on the past domain naturally determines a flow line, which is given the name ``attractor solution''~
	\cite{Heller:2015dha}.

	Next, we would discuss the formal transseries of the Bjorken flow.
	Assuming the flow time to be $\tau$, one can obtain the formal transseries around both $\Sigma_{\rm I(B)}$ and $\Sigma_{\rm I(C)}$.
	For the transseries around $\Sigma_{\rm I(C)}$, we cannot take $1/\tau$ as the expansion parameter.
	This fact can be easily seen from Eq.~(\ref{eq:Bjorken_cw}) by obtaining the leading order behavior of the temperature, $T(\tau) \sim T_0/\tau^{1/3}$ with $T_0 \in {\mathbb R}^+$.
	Thus, in order to construct a transseries that is closed under all operations in the ODEs, one has to take $\tau^{2/3}$ as the expansion parameter.
	Using $z:=\tau^{2/3}$, the ODEs are modified as follows \cite{Behtash:2019txb}
	\be
	&& \frac{d c_{\ell}}{d z} = -  \frac{3 T_0}{2 \theta_0}  \hat{T} c_{\ell}  + \frac{3}{2 z} \left( \frac{2 c_{1}}{15}c_{\ell} + {\frak A}_{\ell} c_{\ell+1} + \bar{\frak B}_{\ell} c_{\ell} + {\frak C}_{\ell} c_{\ell-1} \right) , \\
	&& \hat{T} = \exp \left( - \frac{1}{20} \int \frac{dz}{z} \, c_1    \right),
	\ee
	where the temperature is retrieved by $T(z) = \hat{T}(z) T_0/z^{1/2}$.
	
	The formal transseries for the solution of the Bjorken dynamical system is defined over the ${\mathbb C}$-polynomial ring ${\mathbb C}[ z^{-1}, \frac{\sigma_1 e^{-Sz}}{z^{\beta_1}}, \dots , \frac{\sigma_{L} e^{-Sz}}{z^{\beta_{L}}}]$ with $S=3T_0/(2\theta_0)$ and it turns out to be a divergent series. The so-called non-hydrodynamic modes are associated with the exponential part of the polynomial ring, that clearly shows the nonperturbative nature of the solution. Calculating the Borel transform of the asymptotic power expansion, 
	we can find singular points on the real axis of the Borel plane, verifying the Borel non-summability of the series. 
	On the hand, the formal transseries~\eqref{eq:transsol_c} solution of the Gubser dynamical system is a convergent series with a finite radius of convergence and there are no exponential terms in the polynomial ring as proven in this paper.
	
	In the scenario where $T(\tau)$ is set to zero for every $\tau$, the formal transseries given by Eq.~(\ref{eq:Bjorken_c_T0}) over the polynomial ring ${\mathbb C}[\frac{\sigma_{1}}{z^{\beta_{1}}}, \dots, \frac{\sigma_{L}}{z^{\beta_{L}}}]$ with $z:= \log \tau$.
	Notice that this transseries looks very much like the Gubser solution in that it lacks the exponential profile that existed in the general solution for the case $T\ne0$. Hence, understandably the conclusion on the convergence of the series is again the same as before.
	The polynomial structure ${\mathbb C}[\frac{\sigma_{1}}{z^{\beta_{1}}}, \dots, \frac{\sigma_{L}}{z^{\beta_{L}}}]$ entering a transseries universally appears in a generic autonomous system such as RG flow equation of quantum field theories.
	We invite the interested readers to check Ref.~\cite{Behtash:2019txb} for technical details on the formal transseries solution and flow structure of the Bjorken dynamical system. We summarize the comparison between the Gubser and Bjorken dynamical systems and their transseries solutions in Tab.~\ref{tab:comp_Gub_Bjo}.
	
	\begin{table}[htbp]
		\centering
		\small
		\begin{tabular}{|l||c|c|} \hline
			& \textbf{Gubser} & \textbf{Bjorken} \\ \hline \hline
			Spacetime  & $dS_3 \times {\mathbb R}$ & ${\mathbb R}^{1,3} $ \\
			Symmetry & $SO(3) \times SO(1,1) \times {\mathbb Z}_2$ & $ISO(2) \times SO(1,1) \times {\mathbb Z}_2$ \\
			\hline
			\# of bounded ${\cal S}_a$ ($T \ne 0$) & 6 & 0($\tau^{2/3}$), 2($w$)  \\
			\# of source ($T \ne 0 $)  & 1 & 0($\tau^{2/3}$), 1($w$) \\
			\# of sink ($T \ne 0 $)  & 1 & 1 \\
			$\lim_{\, \rm IR} f/f_{\rm eq}$  & $\ne 1$ (Non-thermal) & $= 1$ (Thermal) \\ \hline
			Expansion param. $z$ & $e^{2 \rho}$ & $\tau^{2/3}, w=\tau T$ \\
			Borel summability  & Convergent series & Non-summable \\
			Formal transseries ($T \ne 0$)  & ${\mathbb C}[ z^{-1}, \sigma_1 z^{\beta_1}, \dots ,\sigma_N z^{\beta_{N}}]$
			& ${\mathbb C}[ z^{-1}, \frac{\sigma_1 e^{-Sz}}{z^{\beta_1}}, \dots , \frac{\sigma_{N}e^{-Sz}}{z^{\beta_{N}}}]$ \\ \hline
		\end{tabular}
		\normalsize
		\caption{Comparison between the Gubser and Bjorken flows. In the first column from the right, $\tau^{2/3}$ and $w$ each stand for a separate candidate for the variable $z$.}
		\label{tab:comp_Gub_Bjo}
	\end{table}
	
	\section{Conclusions and final remarks} \label{sec:conclusions}
	As we have shown in this work, the asymptotic series around the UV/IR fixed points for the Gubser flow is a purely perturbative (or power-law asymptotic) with a finite radius of convergence. This literally means that the asymptotic series converges to the exact solution by adding higher order terms within a suitable truncation scheme. 
	
	In the case of a divergent series, there exists a truncation order that minimizes the error related to the deviance from the exact solution. This error is controlled by the value of the expansion parameter in the transseries solution.
	Therefore, if one keeps adding higher-order terms beyond a certain order, the formal transseries starts to veer off from the exact solution, a fact which was observed in the IR asymptotic expansion of the Bjorken flow~\cite{Behtash:2019txb}. It should be noted that calculating the error systematically is a well-established subject in the superasymptotic and hyperasymptotic approximation methods~\cite{boyd:1999}.
	
	It is also possible to obtain the asymptotic series with the expansion parameter being $\tau:=\tanh \rho \in (-1,+1)$ and setting the initial condition ${\bf c}(\tau=0)=\bar{\bf c}$ where $\bar{\bf c}$ has to lie in the basin of attraction of the IR fixed point (non-thermal equilibrium) at $\tau\rightarrow1$, even though $\bar{\bf c}$ is not a fixed point.
	We then find the evolution equations in terms of $\tau$ from Eqs.~(\ref{eq:ODE_Gub_n0_c}) and (\ref{eq:ODE_Gub_n0_T}) as
	\be
	\frac{ dc_{\ell}}{d \tau} 
	&=& -\left[ \frac{{T}}{\theta_0}c_\ell  +  \left( \frac{2c_1}{15}c_{\ell} +  {\frak A}_{\ell} c_{\ell+1} + \bar{\frak B}_{\ell} c_{\ell}  + {\frak C}_{\ell} c_{\ell-1} \right) \tau \right] \frac{1}{1-\tau^2} , \label{eq:ODE_Gub_n0_c_tau} \\
	\frac{ d {T}}{d \tau} &=& -\frac{{T}}{3} \left( 2 - \frac{c_1}{10} \right) \frac{\tau}{1-\tau^2}, \label{eq:ODE_Gub_n0_T_tau} 
	\ee
	and show that the formal transseries is has a power series form, i.e.
	\be
	\label{eq:powerseries}
	{\bf c}(\tau) = \sum_{k=0}^{+\infty} {\bf u}_{k} \tau^{k}, \qquad \mbox{as \ } \tau \rightarrow 0^{\pm},
	\ee
	where ${\bf u}_{k} \in {\mathbb R}^{L+1}$. The previous expression is, of course, a convergent series. We should notice that this power series can not directly encode the information of the UV/IR fixed points because ${\bf c}(\tau)$ fails to be analytic at $\tau=\pm 1$ (Eqs. \eqref{eq:ODE_Gub_n0_c_tau}-\eqref{eq:ODE_Gub_n0_T_tau} are not obviously regular at these limits). However, one can in principle translate the power series expansion~\eqref{eq:powerseries} into a transseries solution by expanding around either the UV or IR fixed point and perform an analytic continuation. For the Bjorken flow this procedure was previously proposed in Refs.~\cite{Behtash:2019txb,Beuf:2009cx} and was analyzed more recently in~\cite{Kurkela:2019set}. It must be emphasized that here we did not consider analytic continuation of the series~\eqref{eq:powerseries} albeit being a very straightforward calculation. 
	
	We should mention that finding the expansion parameter for a formal transseries solution of a nonlinear ODE
	is not an easy task. The same goes for the Gubser (or Bjorken) dynamical system, in which the expansion parameter is {\it not} an arbitrary variable and should rather be derived from the ODEs themselves. This is not necessarily limited to the transseries and must in fact be taken into account even for the naive asymptotic expansion. For instance, in the Bjorken flow \cite{Behtash:2019txb}, a naive ansatz defined as a power series in terms of $\tau^{-1}$ would not work for solving the dynamical system. Instead, one should use $\tau^{-2/3}$ as the expansion parameter. In general, for a given ODE the appropriate ansatz for the expansion parameter is constructed by ensuring that all the properties of the original theory are kept intact. So it is very risky to
	introduce an arbitrary expansion parameter, as in bookkeeping parameter methods, due to the basic fact the modified ODE might break the well-defined asymptotic properties of the original theory.
	When one assumes the ansatz for the distribution to be of the form $f = f_{\rm eq} + \alpha f^{(1)} + \alpha^2 f^{(2)} + \cdots$ with a bookkeeping parameter $\alpha \in {\mathbb R}^{+}$, it will eventually be necessary to validate whether or not the asymptotics satisfies $|f_{\rm eq}| \gg |f^{(1)}| \gg |f^{(2)}| \gg \cdots$ in the IR limit. This validation test turns out to only pass for the Bjorken flow.
	
In this work, we have established a new set of methods to determine whether a non-equilibrated physical system hydrodynamizes, which is essentially described by \textit{the existence of the slow invariant manifold}. We have shown that in order for hydrodynamization to happen, it is necessary for the dynamics of the system to entail a few degrees of freedom that are slowly relaxing to their equilibrium values. For a system undergoing Bjorken flow described by the RTA Boltzmann equation, the slowest moments $c_{01}$ and $c_{11}$ in the long wavelength limit basically form the \textit{slow invariant manifold}~\cite{Behtash:2019txb} . The Gubser flow does not respect this generic rule since its asymptotic behavior requires an infinite number of moments, none of which are exponentially decaying (nonperturbative). Based on previous studies~\cite{Denicol:2014xca,Denicol:2014tha,Martinez:2017ibh,Behtash:2017wqg} this also explains naturally why the asymptotic gradient expansion breaks down since this perturbative approach relies on being close to the thermal equilibrium. However, the kinetic Bjorken flow fulfills this requirement~\cite{Behtash:2019txb} since its asymptotic IR behavior is controlled by exponentially decaying non-hydrodynamic modes. On the other hand, in Refs.~\cite{Behtash:2018moe,Behtash:2019txb} it was argued that the transport coefficients get dynamically renormalized such that constituent relations can exist in non-equilibrated systems. Our work imposes restrictions on the emergence of non-equilibrium constitutive relations related with the existence of a slow invariant manifold. It would be interesting to explore further these insights and we leave it for future work.
	
As the last comment, we would like to say a few words on the so-called ``attractor solution''. This term has been frequently used in many recent papers on hydrodynamic attractors even though there is no clear definition of the term nor are there any established conditions for its existence.
	At least we can argue that a couple of implicit properties in this regard might come to light by studying these papers which we can relate to: (i) An ``attractor solution'' is a one-dimensional flow line approaching the local equilibrium in the late-time limit; and (ii) (Most of if not) all the flows get attracted to this flow line at large finite flow time irregardless of what the initial condition is.
	The techniques used in this paper and Refs. \cite{Behtash:2018moe, Behtash:2019txb} based on the Morse decomposition and the transseries construction gives the correct mathematical interpretation for these properties and beyond.
	In order to obtain the ``attractor solution'',
	\begin{kotak}
		\begin{itemize}
			\item[(i)] There must exist a pullback attractor ${\cal R}$ with ${\rm Ind}(\Sigma_{\cal R})=1$\footnote{This condition for the pullback attractor can be generally relaxed if the transseries around the fixed point always couples with the integration constants.
				In such cases, one can constrain the fiber space by eliminating the extra dimensions associated to the repelling directions.
			}
			and a forward attractor ${\cal A}$ with ${\rm Ind}(\Sigma_{\cal A})=0$ on the same invariant submanifold (i.e. ${\cal A}<{\cal R}$) where $\Sigma_{{\cal R}({\cal A})}$ is the associated fixed point for ${\cal R}({\cal A})$,
			\item[(ii)] The lowest level transmonomial in the transseries centered at $\Sigma_{\cal A}$ must be decoupled from the integration constants (e.g. the transseries be of the Gevrey-1 class). 
		\end{itemize}
	\end{kotak}

	Only the Bjorken flow (in the $w$-coordinate) satisfies the conditions (i),(ii) all at once. However, the unfortunate fact is that under this time coordinate transformation, the global topology of the flow for the Bjorken dynamical system is changed dramatically in three ways: (1) The most important thing is that the dimension of the fiber space is lowered by one; (2) There is a singularity that appears at $c_{1} = 20$ as seen in Eq.~\eqref{eq:Bjorken_cw}; and (3) The flow at $T=0$ originally defines an autonomous system in \eqref{eq:Bjorken_c_T0}, suggesting that there is a disconnected sector in the configuration space that disappears once the $w$-coordinate is used \cite{Behtash:2019txb}.
	Technically speaking, the mapping $\tau\mapsto w$ for the Bjorken flow cannot continuously deform the topology of the $w$-fiber bundle to that of the $\tau$-fiber bundle.
	Therefore, when $T=0$, namely $w=0$ for any $\tau > 0$, the independent autonomous dynamical structure does not appear when writing the dynamical system in terms of the variable $w$. 
	For $T>0$, focusing only on the subfiber space over the mode configuration $c_{i}$ by excluding the $T$-direction, does mean that $c_{i}$ can stay bounded despite $T$ being divergent in the UV limit. This aspect of the original theory is similar to the flow structure of the transformed system.
	
	It is extremely important to note that if a global attractor is meant by the ``attractor solution'', however, it would basically ensure the existence of a pullback attractor in which case there is a universal attraction by going to the history of the flow; see, for example, \cite{carvalho2012attractors,elia2019existence}.
	We should notice that in the standard textbook definition (e.g. Chap. 3.2 of \cite{Kloeden:2011}), the global attractor possibly does not satisfy both (i) and (ii) above. In such a case, formal transseries around $\Sigma_{\rm U(A)}$ and $\Sigma_{\rm I(A)}$ might have, for example, a form similar to transseries of Gubser flow in the UV and IR limits, respectively. For these reasons, we have rather used the term critical line in \cite{Behtash:2019txb} for putting the emphasis on the fact that this line is just a UV-IR complete flow line and that it lies on the boundary of an invariant submanifold.

	\acknowledgments
	The authors dedicate this article to the memory of Steven Gubser. We also thank S. Grozdanov for illuminating discussions and for reading an earlier version of the draft.
	A. B., M. M., and S. K. are all supported in part by the US Department of Energy Grant No.
	DE-FG02-03ER41260. M. M. is partially supported by the BEST (Beam Energy Scan Theory) DOE Topical Collaboration.
	A. B. was partially supported by the DOE grant DE-SC0013036 and the National Science Foundation under Grant No. NSF
	PHY-1125915.
	
	\bibliography{draft_gubser}
	\bibliographystyle{jhep}
	
\end{document}